\documentclass[journal]{IEEEtran}
\usepackage{amsmath,graphicx,amssymb,multirow,color}
\ifCLASSINFOpdf
\else
\fi
\begin{document}
\title{Audio Source Separation in Reverberant Environments using $\beta$-divergence based Nonnegative Factorization}
\author{Mahmoud Fakhry, Piergiorgio Svaizer, and Maurizio Omologo
       
\thanks{The authors are with the SHINE research unit - Center for Information and Communication Technology (ICT) - Fondazione Bruno Kessler (FBK), Via Sommarive 18, 38123 Trento, Italy.}
}

\markboth{ieee transactions on audio, speech and language processing,~Vol.,~No.}%
{Shell \MakeLowercase{\textit{et al.}}: Bare Demo of IEEEtran.cls for Journals}
\maketitle

\begin{abstract}
In Gaussian model-based multichannel audio source separation, the likelihood of observed mixtures of source signals is parametrized by source spectral variances and by associated spatial covariance matrices. These parameters are estimated by maximizing the likelihood through an Expectation-Maximization algorithm and used to separate the signals by means of multichannel Wiener filtering.

We propose to estimate these parameters by applying nonnegative factorization based on prior information on source variances. In the nonnegative factorization, spectral basis matrices can be defined as the prior information. The matrices can be either extracted or indirectly made available through a redundant library that is trained in advance. In a separate step, applying nonnegative tensor factorization, two algorithms are proposed in order to either extract or detect the basis matrices that best represent the power spectra of the source signals in the observed mixtures. The factorization is achieved by minimizing the $\beta$-divergence through multiplicative update rules. The sparsity of factorization can be controlled by tuning the value of $\beta$.

Experiments show that sparsity, rather than the value assigned to  $\beta$ in the training, is crucial in order to increase the separation performance. The proposed method was evaluated in several mixing conditions. It provides better separation quality with respect to other comparable algorithms.   
\end{abstract}

\begin{IEEEkeywords}
Spectral bases, nonnegative factorization, $\beta$-divergence, extraction, training, detection, estimation, separation. 
\end{IEEEkeywords}

\section{Introduction}
\IEEEPARstart{M}{}ultichannel audio source separation is still a challenging problem. Starting from multiple observations of mixed audio signals generated by different sources, a successful separation system would extract each of the source signals. For instance, in indoor conversations, when 
multiple people are simultaneously talking, each person is considered as a source of an audio signal. If an array of microphones is installed inside the room, each microphone picks up a combination of source spatial images (i.e., copies of the original audio signals filtered through the propagation in the environment). These multiple combinations are considered as observed mixtures of the source signals. A challenging situation arises when the number of microphones is less than the number of sources (under-determined mixing model), and the surrounding mixing environment is reverberant (convolutive mixing model). The problem has been tackled at first with minimal prior knowledge (i.e., blind source separation) \cite{BSS, ICA, cich}. However, an effective solution requires in practice prior information on the source signals \cite{Sun13,Oz12,mahmoud13,mahmoud16} and/or on the mixing environments \cite{FakhryN12,nestafakhry,Du10,Du13}. 

In blind source separation (BSS), many approaches have been proposed in the literature, which work in the time-frequency domain through the short-time Fourier transform (STFT). In frequency-domain independent component analysis (FDICA) \cite{BSS} and clustering \cite{Clust,BM}, the observed mixtures are represented as the product between source signals in the time-frequency domain and complex-valued mixing vectors. In the under-determined mixing model, given an estimate of the mixing vectors and
exploiting the sparsity of audio signals \cite{Bofill, yqli}, the source signals are obtained using binary masking \cite{BM}, soft masking \cite{Clust}, or $l_0$-norm minimization \cite{Lnorm}. 

A time-frequency Gaussian model of audio signals was introduced in \cite{GM}, which allows for separation in the case of underdetermined mixtures. The spatial images of all source signals in the observed mixtures are locally modeled by multivariate Gaussian distributions. The complex covariance matrix of each probability distribution is parametrized by spatial and spectral parameters. Audio propagation channels from a source to microphone positions are statistically represented by a complex spatial covariance matrix, i.e. the spatial component of the model. Moreover, each time-frequency point of the power spectrum of the source signal is described by a scalar source variance, i.e. the spectral component of the model. Assuming that the spatial images of the source signals are independent, the likelihood function of the observed mixtures is a {multivariate Gaussian distribution, which is the product of the distributions of the spatial images of all the source signals}. Based on the parameters of the model estimated by maximizing this likelihood function,  the source signals are obtained by applying multichannel Wiener filtering.

Nonnegative matrix factorization (NMF) can be used to model the source variance as product of two nonnegative vectors \cite{KL, NMTF}, i.e., the power spectrum is decomposed into two nonnegative matrices: a spectral basis matrix containing constitutive parts of the power spectrum, and a coefficient matrix containing time-varying weights. An extension of NMF has been considered by arranging multiple signal observations in a tensor form, where the observations form slices of a $3$-D tensor \cite{NMTF} that is decomposed by means of nonnegative tensor factorization (NTF). The redundancy among the original tensor slices is described by matrices, while the diversity is represented by decomposed tensors. NTF is used for source separation  in an unsupervised way in \cite{virt}, and by incorporating spatial cues as prior knowledge in \cite{Schmi}. The decomposition is achieved by minimizing cost functions, which are error measurement functions. 

Divergence is widely used as cost function to measure the difference between two quantities. For instance, the Kullback-Leibler (KL) divergence \cite{KL} is used to compare two probability distributions. The Itakura-Saito (IS) divergence \cite{IS} is used as a measure of the perceptual differences between spectra. {These divergences are applied to perform NMF for source separation in \cite{kam,kam2}.} The generalized $\beta$-divergence \cite{basu}, used as a cost function for NMF in \cite{beta,Fevot}, encompasses the KL and IS divergences, and it was explored in \cite{Derry} for musical source separation using NTF. The factorization is achieved by minimizing the $\beta$-divergence by means of multiplicative update algorithms (MU).

\subsection{{Overview}}
Audio source separation based on local Gaussian modeling of the mixing process was first proposed in \cite{GM} to study instantaneous linear source separation. In \cite{Oz12,Du10, Du13,Oz10,Ar10}, the model was extended to tackle convolutive source separation. The complex spatial covariance matrix is assumed to be frequency-dependent and time-invariant, and modeled by either a rank-1 mixing model or an unconstrained mixing model. Along with spatial covariance matrix estimation, NMF was used to model the source variance in \cite{Oz12, Oz10, Ar10}. In the case of  semi-supervised source separation for speech enhancement, NMF has been explored using a universal dictionary of pre-trained spectral bases representing multiple source signals, as discussed in \cite{Sun13}. The selection of the optimal bases is done using block sparsity constraints on top of the NMF objective. 

In \cite{mahmoud13},
we used NTF in a semi-supervised scenario to estimate the parameters of the Gaussian model. We also proposed the idea of building a redundant library of spectral basis matrices separately trained on power spectra of multiple sources. Then, the basis matrices that best represent the source signals were detected in the observed mixtures. The detection step is carried out by means of NTF, where multiple observations are used to compute the contribution of each basis vector in the library. The basis matrices with the highest integrated contributions form a subset of basis functions eventually exploited to estimate the parameters.

In this paper, we focus on source separation methods based on local Gaussian modeling of the mixing process. As in the works referenced above, the source variance is modeled using NMF. However, given source-based prior information (e.g., generally deriving from training material that includes single channel/single source speech), multichannel decomposition using NMF/NTF is adopted to estimate the parameters of the model. The advantage introduced by using prior information in the factorization process is that not only the parameters are jointly estimated by factorizing multiple observations, but also unwanted artifacts (e.g., source crosstalk) are avoided. Due to the implicit requirements of nonnegativity (that is applicable only to real numbers), we split the parameters of the model into two sets: a set of nonnegative parameters and another set of complex parameters. NTF is applied to multichannel information for jointly estimating the parameters of the first set. Furthermore, to minimize the scaling ambiguity between the two sets, each parameter of the second set is individually estimated thanks to NMF. In a separate step, we assume that the prior information is either extracted from the observed mixtures, or detected from a redundant library that is trained in advance using NMF. 
Here, with the term \textit{redundant} we denote a library complete enough to contain trained spectral basis matrices suitable to represent any combination of possible sound sources. Both extraction and detection steps are performed by applying NTF to multiple observations. For each task of training and extracting/detecting the prior information as well as for estimating the parameters of the model, a tuning of $\beta$ is required. The choice of this parameter is tightly related to the need of both exploiting the time-frequency sparsity of audio source signals and preserving their basic spectral structure continuity and smoothness.

We highlight that NTF should in principle perform better than NMF when there is redundancy among multiple signal observations, because in NTF multichannel observations are jointly processed in a parallel way. Since the dataset to train the spectral bases consists of several clean speech sentences, their sparse time-frequency representation is not characterized by any usable spatial redundancy. For this reason, in this work we use NMF in training, as NTF would not be effective. This training leads to an efficient representation of the source signals by means of a few spectral basis vectors. On the contrary, for the test data, the same spoken sentences are propagating through several channels. Hence, the spatial redundancy among multiple signal observations is very high. To exploit this redundancy, the multichannel observations are arranged in a 3-D tensor. In this sense, NTF is applied to extract or detect the spectral basis matrices, to estimate the set of nonnegative parameters. After the first estimation, we use the subset of the nonnegative parameters to update the set of complex parameters.         

The rest of the paper is organized as follows. Section II presents the formulation and modeling of the problem. The proposed method is explained in Section III. The extraction, the training and the detection of the source-based prior information are described in Section IV. The experimental analysis is reported in Section V, and Section VI concludes the paper. 

\section{FORMULATION AND MODELING}\label{S:2}
Assume that $N$ sources are observed by an array of $M$ microphones. Using the discrete short time Fourier transform (STFT), each time-frequency point $(\omega,l)$, out of the total number of frequencies $\Omega$ and time frames $L$, of the observed mixtures is represented by a $M \times 1$ vector of complex coefficients $\mathbf{x}(\omega,l)$. The vector can be represented as the combination of $N$ source spatial images $\mathbf{c}_n(\omega,l)$ as 
c
At the microphones, the vector of the spatial images $\mathbf{c}_n(\omega,l)$ of the source signal $s_n(\omega,l)$ is expressed as
\begin{equation}
\mathbf{c}_n(\omega,l)=\mathbf{h}_n(\omega)s_n(\omega,l),
\end{equation}
where $\mathbf{h}_n(\omega)$ is a vector of time-invariant frequency responses. Each element of this vector represents the propagation channel (at the frequency $\omega$) between the \textit{n}-th source and one of the $M$ microphone locations. The vectors $\mathbf{c}_n(\omega,l)$ are assumed to be independent over all the time-frequency points, and they are modeled by a zero-mean multivariate complex Gaussian distribution with a covariance matrix $\mathbf{\Sigma}_{\mathbf{c}_n}(\omega,l)$
\begin{equation} 
\mathbf{c}_n(\omega,l)\sim \mathcal{N}_c(\mathbf{0},\mathbf{\Sigma}_{\mathbf{c}_n}(\omega,l)),
\end{equation}
where $\mathbf{0}$ is a $M \times 1$ vector of zeros. Under the assumption that the source images $\mathbf{c}_n(\omega,l)$, $n=1,..., N$, are statistically independent, the observed mixtures $\mathbf{x}(\omega,l)$ are also modeled by a zero-mean multivariate complex Gaussian distribution with a covariance matrix obtained as
\begin{equation} 
\mathbf{\Sigma}_{\mathbf{x}}(\omega,l)=\sum_{n=1}^N\mathbf{\Sigma}_{\mathbf{c}_n}(\omega,l).
\end{equation}
Maximum likelihood (ML) estimation is shown to be achieved by the minimization of the minus log-likelihood as \cite{Oz10}
\begin{equation}\label{eq:ML}
\xi(\theta)=\sum_{\omega,l}\mathrm{tr}(\mathbf{\Sigma}_{\mathbf{x}}^{-1}(\omega,l)\mathbf{\tilde{R}}_{\mathbf{x}}(\omega,l))+\log|\pi\mathbf{\Sigma}_{\mathbf{x}}(\omega,l)|,
\end{equation}
where $|.|$ denotes the determinant of a square matrix, tr(.) indicates the trace of a matrix, $\theta=\{\mathbf{\Sigma}_{\mathbf{c}_1}(\omega,l), ..., \mathbf{\Sigma}_{\mathbf{c}_N}(\omega,l)\}_{\omega,l}$ is the set of model parameters, and $\mathbf{\tilde{R}}_{\mathbf{x}}(\omega,l)$ is an empirical covariance matrix of the mixtures which can be obtained in a linear form as \cite{Oz12} 
\begin{equation}
{\mathbf{\tilde{R}}}_\mathbf{x}(\omega,l)=\mathbf{x}({\omega},{l})\mathbf{x}^H({\omega},{l}),
\end{equation}
where $.^H$ indicates the conjugate transposition. In the case of a quadratic representation \cite{Oz12}, the matrix is obtained by local averaging over the neighborhood of each time-frequency point as
\begin{equation}\label{eq:ESSIii}
{\mathbf{\tilde{R}}}_\mathbf{x}(\omega,l)=\frac{\sum_{\tilde{\omega},\tilde{l}}\gamma(\tilde{\omega}-\omega,\tilde{l}-l)\mathbf{x}(\tilde{\omega},\tilde{l})\mathbf{x}^H(\tilde{\omega},\tilde{l})}{\sum_{\tilde{\omega},\tilde{l}}\gamma(\tilde{\omega}-\omega,\tilde{l}-l)},
\end{equation}
where $\gamma$ is a bi-dimensional window describing the shape of the neighborhood. As shown in Eq. (\ref{eq:ESSIii}), the quadratic form includes additional information about the local correlation between propagation channels, which often increases the estimation accuracy. The separation is performed by first finding a set $\theta$ that is optimal in the sense of ML. The source spatial images are then obtained in the sense of the minimum mean square error (MMSE) applying multichannel Wiener filtering   
\begin{equation}\label{eq:ESSI}
{\mathbf{\tilde{c}}_n}(\omega,l)= \mathbf{G}_n(\omega,l)\mathbf{x}(\omega,l),
\end{equation}
where the filter gain $\mathbf{G}_n(\omega,l)$ is computed as 
\begin{equation}\label{eq:MWF}
\mathbf{G}_n(\omega,l)= \mathbf{\Sigma}_{\mathbf{c}_n}(\omega,l)\mathbf{\Sigma}_{\mathbf{x}}^{-1}(\omega,l).
\end{equation}

\subsection{{Smooth Wiener filtering}}
The conventional multichannel Wiener filter expressed in (\ref{eq:MWF}) minimizes the mean-square error between the filter input and output signals. Applying extra constraints such as smoothness, or sparseness, on the minimization problem can improve the results. Multichannel spatial smoothing techniques have been proposed in order to widen the spatial response of the filter, so as to reduce possible artifacts spatially close to the target source direction. In this work, spatial smoothing \cite{WF} is used, in which the filter is expressed as 
\begin{equation}\label{eq:smooth}
\mathbf{G}^s_n(\omega,l)= \mathbf{\Sigma}_{\mathbf{c}_n}(\omega,l)[(1-\mu)\mathbf{\Sigma}_{\mathbf{x}}(\omega,l)+\mu \mathbf{\Sigma}_{\mathbf{c}_n}(\omega,l)]^{-1}.
\end{equation}
The smoothness of the resulting filter increases with $\mu$, so that it is equal to the conventional Wiener filter $\mathbf{G}_n(\omega,l)$, for $\mu=0$, or to the identity filter, for $\mu=1$. 

\subsection{Spatial covariance decomposition}
In the spatial covariance decomposition, the covariance matrix of the spatial images of the \textit{n}-th source is modeled as the product between a scalar variance $v_n(\omega,l)$ and a $M\times M$ time-invariant spatial covariance matrix $\mathbf{R}_n(\omega)$. The former encodes the power spectrum of the source signal at each time-frequency point and the latter encodes the spatial information associated with the propagation of the source signal at each frequency 
\begin{equation}\label{eq:SI}
\mathbf{\Sigma}_{\mathbf{c}_n}(\omega,l)= v_n(\omega,l)\mathbf{R}_n(\omega).
\end{equation}
Following such decomposition, the set of the model parameters to estimate is now described by
\begin{equation}
\theta=\{\{v_1(\omega,l), ..., v_N(\omega,l)\}_l,\mathbf{R}_1(\omega) ,..., \mathbf{R}_N(\omega)\}_{\omega}.
\end{equation}
\begin{enumerate}
\item $Spatial~parameters:$ The spatial covariance matrix of each source models the spatial characteristics, such as phase and intensity differences between propagation channels. $\mathbf{R}_n(\omega)$ can be represented by a rank-1 model as $\mathbf{R}_n(\omega)=\mathbf{h}_n(\omega){\mathbf{h}}^H_n(\omega)$. It can also be modeled as an unconstrained matrix, which is considered in this work. As the focus of the paper is spectral modeling, these aspects are not further detailed. 
$\newline$
\item $Spectral~parameters:$ The power spectrum of the source signal $s_n(\omega, l)$ denoted as $\mathbf{V}_n=[v_n(\omega,l)]_{\Omega \times L}$ can be decomposed into the product of two nonnegative matrices, using nonnegative matrix factorization (NMF). Accordingly, the nonnegative source variance $v_n(\omega,l)$ can be represented as the multiplication of two vectors, each with nonnegative entries  
\begin{equation}\label{eq:vnw}
v_n(\omega,l)=\mathbf{u}_n(\omega)\mathbf{{w}}_n(l),
\end{equation}
where $\mathbf{u}^T_n(\omega)$ is a spectral basis column vector of $K$ latent coefficients of the spectral basis matrix $\mathbf{U}_n=[\mathbf{u}_n(\omega)]_{\Omega \times K}$, and $\mathbf{{w}}_n(l)$ is a column vector of $K$ latent coefficients of the time-varying coefficient matrix $\mathbf{{W}}_n=[\mathbf{{w}}_n(l)]_{K\times L}$.
\end{enumerate}

\subsection{Estimation of the model parameters}
The set $\theta$ is estimated by minimizing the criterion in (\ref{eq:ML}), using a generalized expectation maximization algorithm (GEM) \cite{EM} that consists in alternating the following two steps:   
\begin{enumerate}
\item $E~step$: given the observed mixtures $\mathbf{x}(\omega,l)$ and the current estimate of $\theta$, the conditional expectation of the so-called natural statistics is computed.
\newline
\item $M~step$: given the conditional expectation of the natural statistics, the set $\theta$ is updated so as to increase the conditional expectation of the likelihood of the so-called complete data. 
\end{enumerate}
Since the observed data $\mathbf{X}=\{\mathbf{x}(\omega,l)\}_{\omega,l}$ is fully expressed by the unobserved data $\mathbf{C}=\{{\mathbf{{c}}_n}(\omega,l)\}_{n,\omega,l}$ as it is mathematically modeled in (\ref{eq:MM2}), the set of complete data is defined as $\{\mathbf{X}, \mathbf{C}\}$. The natural statistics are defined as the covariance matrix of the conditional probability of the source spatial images \cite{Du10,Ar10}.   

In the $E~step$, given the current estimate of ${\theta}$, the conditional expectation of the natural statistics is computed as \cite{Du10,Ar10}
\begin{equation}\label{eq:Rct}
{\mathbf{\tilde{R}}}_{\mathbf{c}_n}(\omega,l)={\mathbf{\tilde{c}}_n}({\omega},{l}){\mathbf{\tilde{c}}_n}^H({\omega},l)+(\mathbf{I}- \mathbf{G}_n(\omega,l))\mathbf{\Sigma}_{\mathbf{c}_n}(\omega,l),
\end{equation}
where $\mathbf{I}$ is an $M\times M$ identity matrix. ${\mathbf{\tilde{c}}_n}({\omega},{l}){\mathbf{\tilde{c}}_n}^H({\omega},l)$ is a rank-1 empirical covariance matrix in the linear form. 
This matrix can fully describe ${\mathbf{\tilde{R}}}_{\mathbf{c}_n}(\omega,l)$ when $\mathbf{G}_n(\omega,l)=\mathbf{I}$,
i.e., when only one source contributes to the mixture in the time-frequency point $(\omega,l)$. 
Since the audio source signals are known to be sparse, the case $\mathbf{G}_n(\omega,l) \approx \mathbf{I}$ generally holds,and the matrix ${\mathbf{\tilde{R}}}_{\mathbf{c}_n}(\omega,l)$ can be described by the rank-1 matrix ${\mathbf{\tilde{c}}_n}({\omega},{l}){\mathbf{\tilde{c}}_n}^H({\omega},l)$. However, by describing ${\mathbf{\tilde{R}}}_{\mathbf{c}_n}(\omega,l)$ by the rank-1 matrix, we lose
information about the correlation between propagation channels, which reduces the estimation accuracy.

In the $M~step$, the set of model parameters $\theta_n=\{v_n(\omega,l), \mathbf{R}_n(\omega)\}$ is updated by applying the following minimization 
\begin{equation}\label{eq:MINL}
\begin{split}
&\tilde{\theta}_n=\arg\min_{\theta_n}\\&\sum_{\omega,l} \mathrm{tr} (v^{-1}_n(\omega,l)\mathbf{R}^{-1}_n(\omega)\mathbf{\tilde{R}}_{\mathbf{c}_n}(\omega,l))+\log|\pi \mathbf{\Sigma}_{\mathbf{c}_n}(\omega,l)|. 
\end{split}
\end{equation}
Given the matrix ${\mathbf{\tilde{R}}}_{\mathbf{c}_n}(\omega,l)$, the estimation of the parameters 
$v_n(\omega,l)$ and $\mathbf{R}_n(\omega)$ is performed in the sense of ML \cite{Du10,Ar10} by updating one parameter, then the second parameter is updated using the first estimated one. {In \cite{Ar10}, the spatial covariance matrix $\mathbf{R}_n(\omega)$ is estimated using ${\mathbf{\tilde{R}}}_{\mathbf{c}_n}(\omega,l)$ and $v_n(\omega,l)$. The source variance $v_n(\omega,l)$ is updated by estimating the vectors $\mathbf{u}_n(\omega)$ and $\mathbf{{w}}_n(l)$. These vectors are obtained by applying the KL divergence, given the spatial covariance matrix $\mathbf{R}_n(\omega)$, and a covariance matrix ${\mathbf{\tilde{R}}}_{\mathbf{c}_{n,k}}(\omega,l)$ describing the $k-th$ component $u_n(\omega,k){w}_n(k,l)$}. The main limitation of this method is that the estimation error is accumulated from one parameter to the second, and from one iteration to another.  Furthermore, the vectors $\mathbf{u}_n(\omega)$ and $\mathbf{{w}}_n(l)$ are estimated by combining multiple observations in a single one. As a result, the redundancy between these observations is not well exploited. Moreover, the KL divergence does not constrain the sparsity of factorization.

\section{METHOD}\label{S:3}
Let us assume that we have a corrupted version $\hat{v}_n(\omega,l)$ of the source variance ${v}_n(\omega,l)$, where the distortion is caused by other source signals, multipath propagation, and 
environmental noise. The corrupted source variance can be approximated as 
\begin{equation}
\hat{v}_n(\omega,l)=\mathbf{u}_n(\omega)\mathbf{\hat{w}}_n(l)\approx \mathbf{u}_n(\omega)\mathbf{{w}}_n(l)+\mathbf{\epsilon}(\omega,l),
\end{equation}
where $\mathbf{\hat{w}}_n(l)$ indicates the time-varying vector of $\hat{v}_n(\omega,l)$ in the factorization domain, and $\mathbf{\epsilon}(\omega,l)$ includes the distortion error as well as the factorization error in the estimation of ${v}_n(\omega,l)$. If the basis vector $\mathbf{u}_n(\omega)$ well describes the true source variance ${v}_n(\omega,l)$, a properly good estimation of the vector $\mathbf{{w}}_n(l)$ is obtained. Following this model, the vector $\mathbf{{w}}_n(l)$ can be obtained by applying an efficient factorization algorithm, based on $\mathbf{u}_n(\omega)$ and the observed $\hat{v}_n(\omega,l)$.
The factorization is achieved using divergences, which are distance-like functions that measure the separation between two elements $a$ and $b$. The generalized $\beta$-divergence $d_{\beta}(a|b)$ is expressed as \cite{Fevot} 
\begin{equation}\label{eq:betadiv}
d_{\beta}(a|b)=
\begin{cases} 
\frac{a}{b}-\log(\frac{a}{b})-1, & \beta=0\\
a\log (\frac{a}{b})+b-a, & \beta=1\\
\frac{a^{\beta}+(\beta-1)b^{\beta}-\beta ab^{\beta-1}}{\beta(\beta-1)}, & otherwise
\end{cases}
\end{equation}
\begin{figure}
\centering
\includegraphics[scale=0.5]{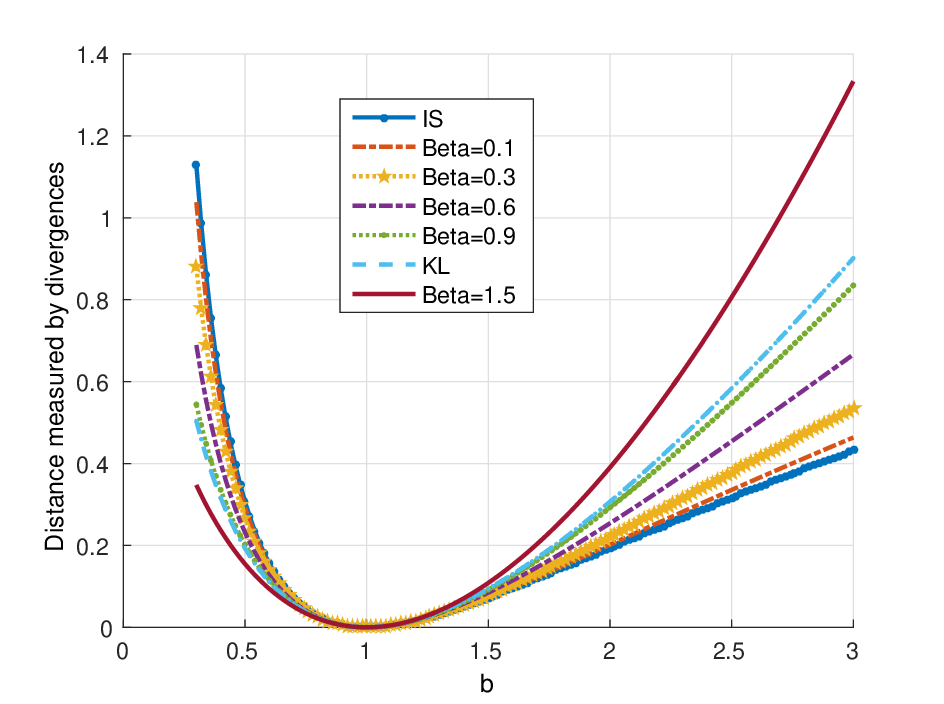}
\caption{Distance measured by different definitions of divergence, with $a=1$.}
\label{fig:div}
\end{figure}
\begin{figure*}[ht]
\centering
\includegraphics[width=18cm,height=3.7cm]{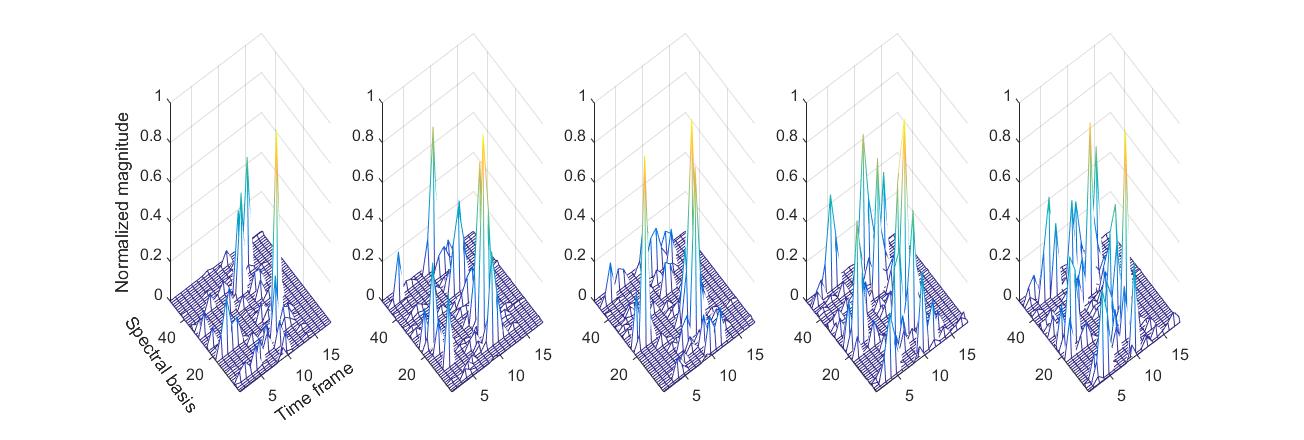}
\caption{Examples of controlling the sparsity of $\mathbf{W}_n$ of the corrupted power spectrum by selecting the value of $\beta$, from the left to the right, respectively, original $\mathbf{W}_n$ (training) with $\beta=0.9$, estimated with $\beta=0.1$, estimated with $\beta=0.3$, estimated with $\beta=0.6$, and estimated with $\beta=0.9$.}
\label{fig:facto2}
\end{figure*}  
\begin{figure*}[ht]
\centering
\includegraphics[width=16cm,height=3.8cm]{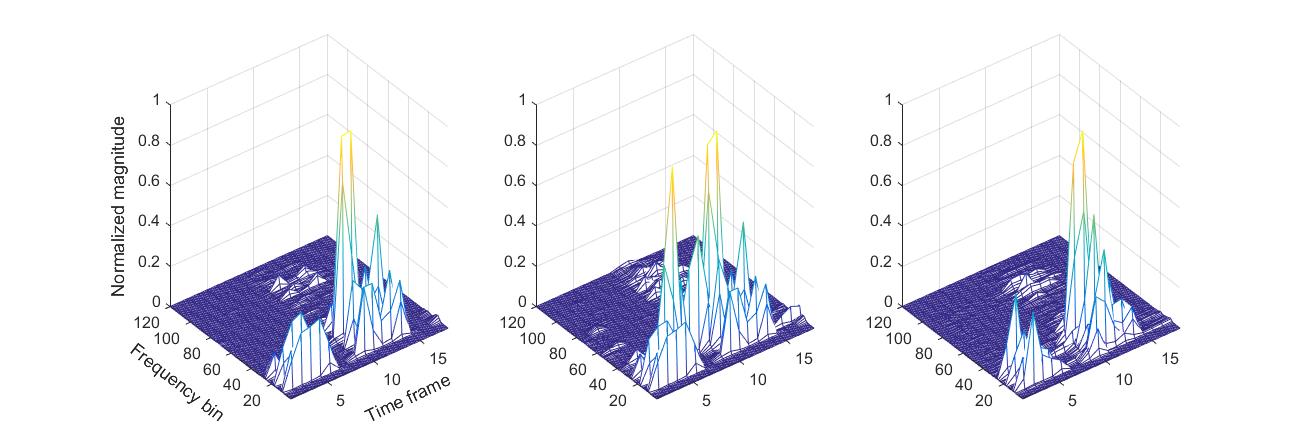}
\caption{Normalized power spectra of true, corrupted, and reconstructed signals, from the left to the right respectively.}
\label{fig:facto}
\end{figure*} 
For $\beta=1$, $d_{\beta}(a|b)$ is the KL divergence, while it is the IS divergence for $\beta=0$. 
{Fig. \ref{fig:div} shows $\beta$-divergence based distances as a function of $b$ with $a=1$, which were obtained by applying different values of $\beta$.

In the context addressed by our work, $b$ is larger than $a$ most of the time.
Consequently, the IS divergence provides a distance that leads to a higher sparsity constraint. 
This sparseness effect can be reduced by increasing the value of $\beta$. Fig. \ref{fig:facto2} and \ref{fig:facto} demonstrate this fact, presenting an example of application of the $\beta$-divergence to minimize the influence of an interference added to the original signal.

We trained $50$ vectors $\mathbf{u}_n(\omega)$ using the power spectrum of a male speech signal, applying the $\beta$-divergence with $\beta=0.9$ and using the multiplicative update rule (MU). The male speech signal was linearly mixed with a second female speech signal using a mixing vector with coefficients $[1,~0.7]$. Observing the power spectrum of the mixture, and given the vectors $\mathbf{u}_n(\omega)$, we want to reconstruct the power spectrum of the male speech signal by estimating its coefficient vectors $\mathbf{{w}}_n(l)$. Applying the $\beta$-divergence in a semi-supervised scenario, (i.e., with only one of the two sources seen during the training)
the sparseness of vectors $\mathbf{{w}}_n(l)$ depends on the value of $\beta$. Fig. \ref{fig:facto2} shows that the sparsity of the estimated vectors $\mathbf{{w}}_n(l)$ can be controlled by tuning this parameter. Accordingly, the value of $\beta$ can be set to minimize the impact of residual artifacts in signal observations, which results in a better estimation. {Among the tested values of $\beta$, Fig. \ref{fig:facto} shows the reconstructed power spectrum with $\beta=0.3$}.  

Building on this idea, the minimization problem in (\ref{eq:MINL}) is reformulated by using the $\beta$-divergence, as done in the semi-supervised case, leading to
\begin{equation}\label{theta_mia}
\tilde{\theta}_n=\arg\min_{\theta_n} \sum_{m_1,m_2}\sum_{\omega,l}d_{\beta}[{\tilde{r}}^{m_1m_2}_{\mathbf{c}_n}(\omega,l)|v_n(\omega,l){r}^{m_1m_2}_n(\omega)],
\end{equation}
where ${\tilde{r}}^{m_1m_2}_{\mathbf{c}_n}(\omega,l)$ and ${r}^{m_1m_2}_n(\omega)$ are the ($m_1,m_2$) coefficients of the matrices $\mathbf{\tilde{R}}_{\mathbf{c}_n}(\omega,l)$ and $\mathbf{R}_n(\omega)$, respectively, and $m_1,m_2=1,..., M$. 
Let us replace the linear form by the quadratic form in order to include information about the neighborhood in the computation of the covariance matrix. Then, we propose to compute
${\mathbf{\tilde{R}}}_{\mathbf{c}_n}(\omega,l)$ by weighted averaging, using smoothing spectral-temporal windowing as in (\ref{eq:ESSIii}) instead of the single point computation as in (\ref{eq:Rct}),
\begin{equation}\label{Rc}
{\mathbf{\tilde{R}}}_{\mathbf{c}_n}(\omega,l)={\mathbf{\hat{R}}}_{\mathbf{c}_n}(\omega,l)+(\mathbf{I}-\mathbf{G}_n(\omega,l))\mathbf{\Sigma}_{\mathbf{c}_n}(\omega,l),
\end{equation}
where ${\mathbf{\hat{R}}}_{\mathbf{c}_n}(\omega,l)$ is computed as in (\ref{eq:ESSIii}), replacing $\mathbf{x}(\omega,l)$ by ${\mathbf{\tilde{c}}_n}(\omega,l)$. 
Substituting
the source variance $v_n(\omega,l)$ in (\ref{theta_mia})
by its decomposition, the parameters are estimated by solving the following minimization problem  
\begin{equation}\label{eq:MLNMF}
\begin{split}
&\tilde{\theta}_n=\arg\min_{\theta_n}\\&\sum_{m_1,m_2} \sum_{\omega,l}d_{\beta}[{\tilde{r}}^{m_1m_2}_{\mathbf{c}_n}(\omega,l)|\mathbf{u}_n(\omega)\mathbf{{w}}_n(l){r}^{m_1m_2}_n(\omega)].
\end{split}
\end{equation}
In order to reduce the degree of freedom of the decomposition, we can assume that one or more of the coefficients/vectors are known (prior information). Since the focus of the paper is on exploiting source-based information, we assume that the frequency dependent spectral basis vector $\mathbf{u}_n(\omega)$ is pre-known, and that the task is to estimate $\mathbf{{w}}_n(l)$ and ${r}^{m_1m_2}_n(\omega)$ by decomposition. Hence the set of parameters is redefined again as ${\theta}_n=\{\mathbf{{w}}_n(l),\mathbf{R}_n(\omega)\}$. Fig. \ref{fig:Method} shows the processing flow of the proposed method. On the other hand, since both $\mathbf{\tilde{R}}_{\mathbf{c}_n}(\omega,l)$ and $\mathbf{R}_n(\omega)$ are complex covariance matrices, the nonnegativity constraint of the factorization can only be applied to their diagonal coefficients. For this reason, we split ${\theta}_n$ into two subsets, a set of nonnegative parameters ${\theta}^{diag}_n=\{\mathbf{{w}}_n(l), {r}^{m_1m_2}_n(\omega)\}$, $m_1=m_2$, and another set of complex parameters ${\theta}^{off}_n=\{{r}^{m_1m_2}_n(\omega)\}$, $m_1\neq m_2$, where ${\theta}_n={\theta}^{diag}_n\cup{\theta}^{off}_n$. 
\begin{figure}
\centering
\includegraphics[width=8.8cm,height=7.9cm]{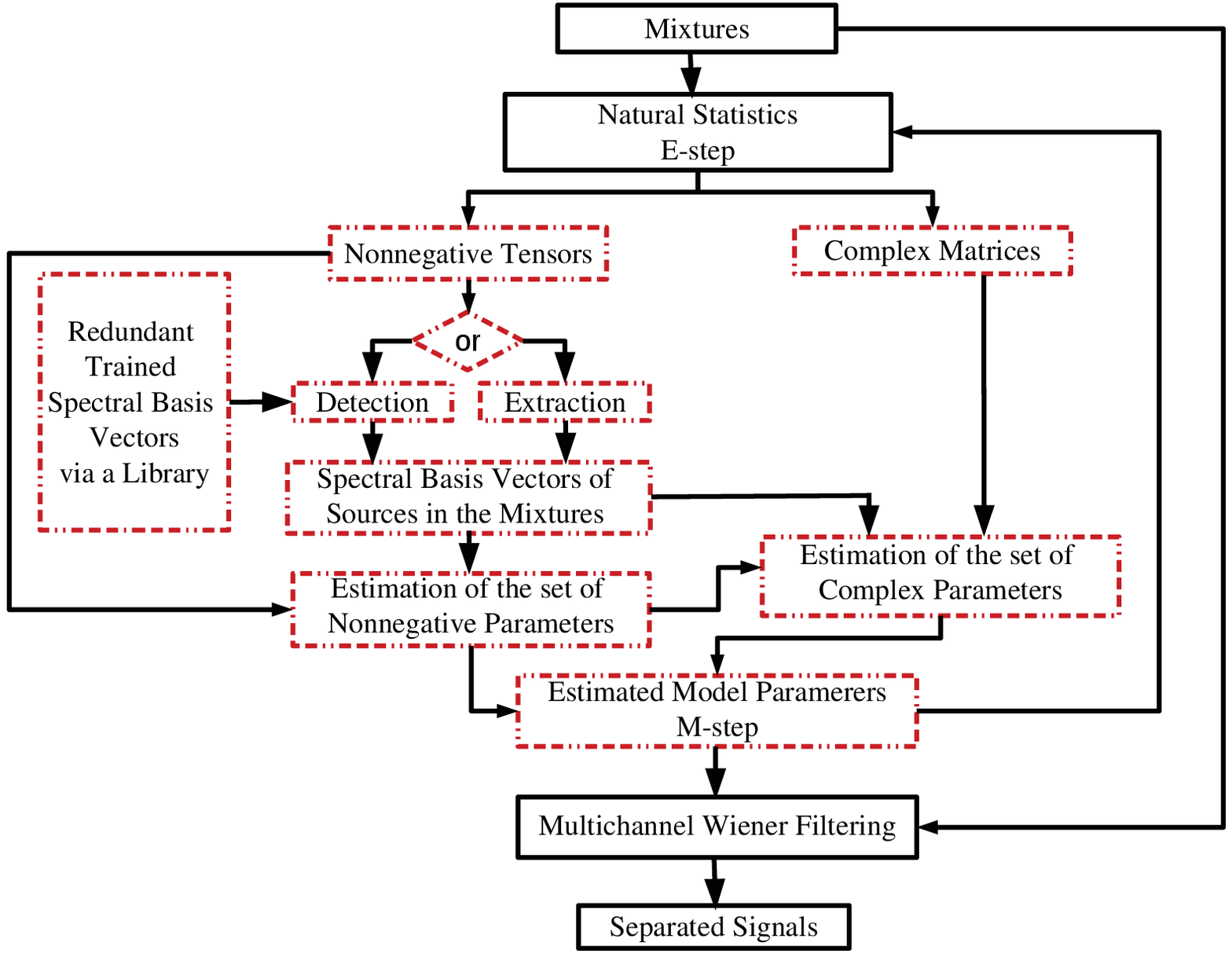}
\caption{Flowchart of the proposed method}
\label{fig:Method}
\end{figure}
The diagonal nonnegative entries of $\mathbf{\tilde{R}}_{\mathbf{c}_n}(\omega,l)$ are multiple observations, where each of them can be approximately represented by the product between the source variance and the propagation channel intensity.
The source variance is fixed from one observation to another, while, the channel intensity changes. 
Hence, these diagonal observations are used to jointly estimate the parameters of the subset
${\theta}^{diag}_n$ by applying NTF.
Moreover, each off-diagonal coefficient of $\mathbf{\tilde{R}}_{\mathbf{c}_n}(\omega,l)$ is used to estimate a corresponding parameter of the subset ${\theta}^{off}_n$ by using NMF. As detailed in the following, we update one complex parameter ${r}^{m_1m_2}_n(\omega)$, $m_1\neq m_2$ 
while the other two nonnegative parameters, $\mathbf{u}_n(\omega)$ and $\mathbf{{w}}_n(l)$, are kept fixed.
   
\subsection{Tensor/matrix representations of $\mathbf{\tilde{R}}_{\mathbf{c}_n}(\omega,l)$ and $\mathbf{R}_n(\omega)$}\label{tensor}
A tensor $\mathbf{\tilde{V}}^M_{\mathbf{c}_n}$ of size $\Omega \times L \times M$ can be built from the diagonal elements of the matrix $\mathbf{\tilde{R}}_{\mathbf{c}_n}(\omega,l)$. The \textit{m}-th diagonal element of this matrix is denoted as ${\tilde{r}^m}_{\mathbf{c}_n}(\omega,l)$, and, accordingly, the \textit{m}-th slice of the tensor is defined as $\mathbf{\tilde{V}}^m_{\mathbf{c}_n}=[{\tilde{r}^m}_{\mathbf{c}_n}(\omega,l)]_{\Omega\times L}$. The same is done in order to define a tensor with diagonal slices of spatial information, $\mathbf{V}^m_{\mathbf{R}_n}=diag~[r^m_n(\omega)]_{\Omega}$, where $r^m_n(\omega)$ indicates the \textit{m}-th diagonal element of $\mathbf{R}_n(\omega)$. Based on the built tensors and the matrices $\mathbf{U}_n$ and $\mathbf{{W}}_n$, the minimization problem to derive the subset of nonnegative parameters in (\ref{eq:MLNMF}) becomes
\begin{equation}\label{eq:Diag}
\tilde{\theta}^{diag}_n=\arg\min_{\theta^{diag}_n}\sum_{m} \sum_{\omega,l}d_{\beta}[\mathbf{\tilde{V}}^m_{\mathbf{c}_n}|\mathbf{V}^m_{\mathbf{R}_n}\mathbf{U}_n\mathbf{{W}}_n],
\end{equation}
where the subset of parameters to estimate is defined as ${\theta}^{diag}_n=\{\mathbf{{W}}_n, \mathbf{V}^m_{\mathbf{R}_n}\}$. 

To complete the formulation of the problem we need to address the off-diagonal coefficients of
$\mathbf{\tilde{R}}_{\mathbf{c}_n}(\omega,l)$ and $\mathbf{R}_n(\omega)$.
For each of the two matrices, the coefficients are complex conjugate of each other, centred around their diagonal coefficients, i.e. the ($m_1,m_2$) coefficient is the complex conjugate of the ($m_2,m_1$) coefficient. Consequently, only half of the coefficients in ${\theta}^{off}_n$ must be estimated.
The ($m_1,m_2$) complex coefficients of $\mathbf{R}_n(\omega)$ are arranged in a diagonal matrix as $\mathbf{V}^{m_1m_2}_{\mathbf{R}_n}=diag~[r^{m_1m_2}_n(\omega)]_{\Omega}$, which covers the whole frequency range. From the ($m_1,m_2$) complex coefficients of $\mathbf{\tilde{R}}_{\mathbf{c}_n}(\omega,l)$, a matrix  $\mathbf{\tilde{V}}^{m_1m_2}_{\mathbf{c}_n}=[{\tilde{r}^{m_1m_2}}_{\mathbf{c}_n}(\omega,l)]_{\Omega\times L}$ is derived. Starting from both $\mathbf{U}_n$ and the $\mathbf{{W}}_n$ estimated by the previous minimization step, the new optimization problem is defined as      
\begin{equation}\label{eq:offD}
\tilde{\theta}^{off}_n=\arg\min_{\theta^{off}_n}\sum_{\omega,l,m_1 \neq m_2}d_{\beta}[\mathbf{\tilde{V}}^{m_1m_2}_{\mathbf{c}_n}|\mathbf{V}^{m_1m_2}_{\mathbf{R}_n}\mathbf{U}_n\mathbf{{W}}_n],
\end{equation}
where the subset of the complex off-diagonal parameters to estimate is represented by
${\theta}^{off}_n=\{\mathbf{V}^{m_1m_2}_{\mathbf{R}_n}\}$, $m_1\neq m_2$.

\subsection{Tensor/matrix update}\label{fact}
Tensor/matrix factorization (NTF/NMF) is achieved by minimizing the $\beta$-divergence. The MU algorithm is applied in order to solve the minimization problems in (\ref{eq:Diag}) and (\ref{eq:offD}). The rule consists of updating each scalar parameter in ${\theta}_n$ by multiplying its value at a previous iteration by the ratio of the negative and positive parts of the derivative of the $\beta$-divergence with respect to the parameter. 

NTF is used to derive the first subset of parameters ${\theta}^{diag}_n$, based on the tensor of multichannel observations $\mathbf{\tilde{V}}^M_{\mathbf{c}_n}$, 
\textit{which accounts for the spatial redundancy among the observations}. 
To estimate both the tensor slice $\mathbf{V}^m_{\mathbf{R}_n}$ and the matrix $\mathbf{W}_n$ by minimizing the $\beta$-divergence, the MU algorithm operates as follows (see more details in the Appendix \ref{appendix})  
\begin{equation}\label{eq:Wn}
\mathbf{W}_n\leftarrow \mathbf{W}_n\circ\frac{\sum_{m}(\mathbf{V}^m_{\mathbf{R}_n}\mathbf{U}_n)^T[\mathbf{\tilde{V}}^m_{\mathbf{c}_n}\circ(\mathbf{V}^m_{\mathbf{R}_n}\mathbf{U}_n\mathbf{W}_n)^{\beta^s-2}]}{\sum_{m}(\mathbf{V}^m_{\mathbf{R}_n}\mathbf{U}_n)^T (\mathbf{V}^m_{\mathbf{R}_n}\mathbf{U}_n\mathbf{W}_n)^{\beta^s-1}},\\
\end{equation}
\begin{equation}\label{eq:Vr}
\mathbf{V}^m_{\mathbf{R}_n}\leftarrow \mathbf{V}^m_{\mathbf{R}_n}\circ\frac{[\mathbf{\tilde{V}}^m_{\mathbf{c}_n}\circ(\mathbf{V}^m_{\mathbf{R}_n}\mathbf{U}_n\mathbf{W}_n)^{\beta^s-2}](\mathbf{U}_n\mathbf{{W}}_n)^{T}}{(\mathbf{V}^m_{\mathbf{R}_n}\mathbf{U}_n\mathbf{W}_n)^{\beta^s-1}(\mathbf{U}_n\mathbf{{W}}_n)^T},\\
\end{equation}
where $\circ$ indicates element-wise multiplication, $\beta^s$ denotes the value of $\beta$ used for estimating the parameters. The division is an element-wise operation. 
From this formulation it is evident that in NTF each slice $\mathbf{V}^m_{\mathbf{R}_n}$ is independently updated. However, the matrix $\mathbf{W}_n$ is updated at once, which allows to exploit the multichannel spatial redundancy. 

As the matrix $\mathbf{W}_n$ is estimated, the second subset ${\theta}^{off}_n$ is derived by using the matrix $\mathbf{U}_n$ in the same way the spatial information $\mathbf{V}^m_{\mathbf{R}_n}$ was estimated in the previous step, but using different observations, such as
\begin{equation}\label{eq:Rmm}
\begin{split}
&\mathbf{V}^{m_1m_2}_{\mathbf{R}_n}\leftarrow\\& \mathbf{V}^{m_1m_2}_{\mathbf{R}_n}\circ\frac{[\mathbf{\tilde{V}}^{m_1m_2}_{\mathbf{c}_n}\circ (\mathbf{V}^{m_1m_2}_{\mathbf{R}_n}\mathbf{U}_n\mathbf{W}_n)^{\beta^s-2}](\mathbf{U}_n\mathbf{{W}}_n)^{T}}{(\mathbf{V}^{m_1m_2}_{\mathbf{R}_n}\mathbf{U}_n\mathbf{W}_n)^{\beta^s-1}(\mathbf{U}_n\mathbf{{W}}_n)^T}.
\end{split}
\end{equation}
In this step, the complex coefficients of $\mathbf{R}_n(\omega)$ are updated based on complex coefficients of $\mathbf{\tilde{R}}_{\mathbf{c}_n}(\omega,l)$, on estimated nonnegative $\mathbf{{W}}_n$, as well as on prior information $\mathbf{U}_n$. In practice, we do this to keep the scaling in equations (\ref{eq:Vr}) and (\ref{eq:Rmm}) unchanged, otherwise a scaling ambiguity would be generated between the estimated diagonal and off-diagonal coefficients of $\mathbf{R}_n(\omega)$. 

\subsection{Initialization}
The tensor slice $\mathbf{V}^m_{\mathbf{R}_n}$ and the matrix $\mathbf{V}^{m_1m_2}_{\mathbf{R}_n}$ are initialized as identity matrices at the estimation step, in order to maintain the diagonal representation. Moreover, the matrix $\mathbf{{W}}_n$ is initialized by random values larger than zero. The source spatial images $\mathbf{\tilde{c}}_n(\omega,l)$ are initialized by binary clustering the time-frequency points of the observed mixture $\mathbf{x}(\omega,l)$. The time difference of arrival (TDOA) of each source signal is estimated as in \cite{GCC}. Given the estimated TDOAs of the multiple sources, the time-frequency points of the observed mixtures $\mathbf{x}(\omega,l)$ are grouped into multiple clusters, each one corresponding to a source signal. This step is performed by minimizing the error between steering vectors of the estimated TDOAs and phase differences of the observed $\mathbf{x}(\omega,l)$.

\section{SOURCE-BASED PRIOR INFORMATION}
As described in the previous sections,
the spectral basis matrix $\mathbf{U}_n$ of the \textit{n}-th source is a pre-known information that can be extracted in a separate step, or assumed to be available in advance, for instance based on pre-training. In the latter case, the matrix is available either directly or indirectly. If indirectly available, we assume to have a redundant library of source spectral basis matrices among which the matrices that best match the observed mixtures are to be detected. As shown later, the proposed method for the extraction of $\mathbf{{U}}_n$ is effective for blind source separation (BSS) in environments with a low reverberation. However, using the pre-trained matrix $\mathbf{U}_n$ the separation performance also increases under higher reverberant conditions.

\subsection{Extraction of the prior information}\label{sec:ESU}
For an automatic unsupervised extraction of the matrix $\mathbf{U}_n$ from the tensor of observations $\mathbf{\tilde{V}}^M_{\mathbf{c}_n}$, $\beta$-divergence minimization problem is formulated as
\begin{equation}\label{EX}
\mathbf{\tilde{U}}_n=\arg\min_{\mathbf{U}_n}\sum_{m} \sum_{\omega,l}d_{\beta}[\mathbf{\tilde{V}}^m_{\mathbf{c}_n}|\mathbf{U}_n\mathbf{\tilde{W}}^m_n],
\end{equation}
where $\mathbf{\tilde{U}}_n$ is an estimation of the spectral basis matrix of the \textit{n}-th source. $\mathbf{\tilde{W}}^m_n$ denotes the \textit{m}-th slice of a time-varying coefficient tensor $\mathbf{\tilde{W}}^M_n$, where each slice is associated with the corresponding \textit{m}-th slice of the tensor $\mathbf{\tilde{V}}^M_{\mathbf{c}_n}$. In this formulation, the value of $\beta$ can be tuned to control a factorization trade-off between sparsity and continuity in the spectral structure of the extracted spectral basis vectors. Moreover, we can minimize the impact of both reverberation and artifacts due to other source signals. The larger the value of $\beta$, the smaller the weight applied to low energy points of $\mathbf{\tilde{V}}^M_{\mathbf{c}_n}$. On the other hand, a small positive value of $\beta$ increases the contribution of low energy points. Accordingly, assigning a suitable value to $\beta$, only high energy points associated with the \textit{n}-th source are used to train the basis vectors of $\mathbf{\tilde{U}}_n$, while other points associated with artifacts and reverberation are ignored.} The diversity and the time-indeterminacy among the slices of the tensor $\mathbf{\tilde{V}}^M_{\mathbf{c}_n}$ are also {transferred} to the tensor $\mathbf{\tilde{W}}^M_n$, while the redundant information is kept in $\mathbf{\tilde{U}}_n$.  The matrix is extracted by applying the MU to minimize the $\beta$-divergence in (\ref{EX}) as (see the {Appendix \ref{appendix})
\begin{equation}
\mathbf{\tilde{U}}_n\leftarrow \mathbf{\tilde{U}}_n\circ\frac{\sum_m[\mathbf{\tilde{V}}^m_{\mathbf{c}_n}\circ(\mathbf{\tilde{U}}_n\mathbf{\tilde{W}}^m_n)^{\beta^e-2}](\mathbf{\tilde{W}}^m_n)^{T}}{\sum_m(\mathbf{\tilde{U}}_n\mathbf{\tilde{W}}^m_n)^{\beta^e-1}(\mathbf{\tilde{W}}^m_n)^{T}},\\
\end{equation}
\begin{equation}
\mathbf{\tilde{W}}^m_n\leftarrow \mathbf{\tilde{W}}^m_n\circ\frac{\mathbf{\tilde{U}}_n^T[\mathbf{\tilde{V}}^m_{\mathbf{c}_n}\circ(\mathbf{\tilde{U}}_n\mathbf{\tilde{W}}^m_n)^{\beta^e-2}]}{\mathbf{\tilde{U}}_n^T (\mathbf{\tilde{U}}_n\mathbf{\tilde{W}}^m_n)^{\beta^e-1}},
\end{equation}
{where}
$\beta^e$ denotes the value of $\beta$ used for extracting the matrix $\mathbf{\tilde{U}}_n$. While iterating these steps each column of $\mathbf{\tilde{U}}_n$ is normalized to sum up to one after each iteration. The tensor $\mathbf{\tilde{W}}^M_n$ is not needed anymore. 

\subsection{Training of the prior information}\label{library}
The matrix $\mathbf{U}_n$ can be pre-trained by applying NMF to a separate set of training audio signals. To this purpose, clean speech signals of the \textit{n}-th source are processed to concatenate power spectra of several utterances into a matrix $\mathbf{V}^t_n$. The matrix $\mathbf{U}_n$ is trained by factorizing $\mathbf{V}^t_n$ into the multiplication of two matrices $\mathbf{U}_n$ and $\mathbf{{W}}^t_n$ as 
{(see the Appendix \ref{appendix})}
\begin{equation}\label{eq:training1}
\mathbf{{U}}_n\leftarrow \mathbf{{U}}_n\circ\frac{[\mathbf{{V}}^t_n\circ(\mathbf{{U}}_n\mathbf{{W}}^t_n)^{\beta^t-2}](\mathbf{{W}}^t_n)^{T}}{(\mathbf{{U}}_n\mathbf{{W}}^t_n)^{\beta^t-1}(\mathbf{{W}}^t_n)^{T}},\\
\end{equation}
\begin{equation}\label{eq:training2}
\mathbf{{W}}^t_n\leftarrow \mathbf{{W}}^t_n\circ\frac{\mathbf{{U}}_n^T[\mathbf{{V}}^t_n\circ(\mathbf{{U}}_n\mathbf{{W}}^t_n)^{\beta^t-2}]}{\mathbf{{U}}_n^T (\mathbf{{U}}_n\mathbf{{W}}^t_n)^{\beta^t-1}},
\end{equation}
where $\beta^t$ denotes the value of $\beta$ used for training the matrix $\mathbf{{U}}_n$. The training coefficient matrix $\mathbf{{W}}^t_n$ is not needed anymore. Assuming that the spatial locations of 
the speakers are labelled for each source, the spectral basis matrix $\mathbf{U}_n$ can be predefined and fixed for each label, which means that the source order must be known in advance. However, a redundant library of spectral basis matrices of all possible sources can be built, in order to increase the flexibility of the algorithm. Then, we detect the basis matrices that match the source signals in the observed mixture. The spectral basis matrices are trained and sequentially arranged side by side in order to constitute the library $\mathbf{U}_{lib}$, with
\begin{equation}\label{LIB}
\mathbf{U}_{lib}=[\mathbf{U}_1| \mathbf{U}_2| \cdots| \mathbf{U}_Z],
\end{equation}
for a number $Z$ of training source signals, where $Z>N$.

\subsection{Detection of the matched prior information}\label{detection}
The tensor $\mathbf{\tilde{V}}^M_{\mathbf{c}_n}$ approximately represents the source power spectrum $\mathbf{V}_n$ weighted by the time-invariant intensity of propagation channels. The spectrum can be decomposed as $\mathbf{V}_n=\mathbf{U}_n\mathbf{W}_n$ by using NMF. Assuming that the matrix $\mathbf{U}_n$ is included in the library $\mathbf{U}_{lib}$, the factorization can be expanded as $\mathbf{V}_n=\mathbf{U}_{lib}\mathbf{D}_{lib}\mathbf{W}_{lib}$, involving a diagonal matrix $\mathbf{D}_{lib}$ whose diagonal coefficients define the contribution of each spectral basis vector in the library $\mathbf{U}_{lib}$. The coefficients of $\mathbf{D}_{lib}$ that are associated with the matrix $\mathbf{U}_n$, will have the largest values over all the other coefficients. Accordingly, the matrix $\mathbf{U}_n$ that best represents the spectrum $\mathbf{V}_n$ can be identified in the library $\mathbf{U}_{lib}$ by observing the diagonal coefficients of the matrix $\mathbf{D}_{lib}$. Though the idea can be extended to multiple observations, it is worth mentioning that $\mathbf{\tilde{V}}^M_{\mathbf{c}_n}$ is a tensor of weighted and corrupted power spectra. The weights are channel intensities, and the corruption is due to residual from other sources, reverberation and environmental noise. As a result, an efficient factorization algorithm is required in order to compensate for the impact of the weights and the corruption. To derive the matched matrices form $\mathbf{\tilde{V}}^M_{\mathbf{c}_n}$, the $\beta$-divergence minimization problem is formulated as
\begin{equation}\label{det}
\mathbf{D}^M_{lib}=\arg\min_{\mathbf{D}^M_{lib}}\sum_{m} \sum_{\omega,l}d_{\beta}[\mathbf{\tilde{V}}^m_{\mathbf{c}_n}|\mathbf{U}_{lib}\mathbf{D}^m_{lib}\mathbf{W}_{lib}],
\end{equation}
where $\mathbf{D}^M_{lib}$ is a tensor with diagonal slices. The \textit{m}-th slice of the tensor and the coefficient matrix $\mathbf{W}_{lib}$ can be computed by applying MU to minimize the $\beta$-divergence in (\ref{det}) as follows 
{(see the Appendix \ref{appendix})} 
\begin{equation}\label{eq:detect1} 
\mathbf{D}^m_{lib}\leftarrow \mathbf{D}^m_{lib}\circ\frac{\mathbf{U}^T_{lib}[\mathbf{\tilde{V}}^m_{\mathbf{c}_n}\circ(\mathbf{U}_{lib} \mathbf{D}^m_{lib} \mathbf{W}_{lib})^{\beta^d-2}]\mathbf{{W}}_{lib}^{T}}{\mathbf{U}^T_n(\mathbf{U}_{lib} \mathbf{D}^m_{lib} \mathbf{W}_{lib})^{\beta^d-1}\mathbf{{W}}_{lib}^T},\\
\end{equation}
\begin{equation}\label{eq:detect2}
\begin{split}
&\mathbf{W}_{lib}\leftarrow\\& \mathbf{W}_{lib}\circ\frac{\sum_{m}(\mathbf{U}_{lib}\mathbf{D}^m_{lib})^T[\mathbf{\tilde{V}}^m_{\mathbf{c}_n}\circ(\mathbf{U}_{lib} \mathbf{D}^m_{lib} \mathbf{W}_{lib})^{\beta^d-2}]}{\sum_{m}(\mathbf{U}_{lib}\mathbf{D}^m_{lib})^T (\mathbf{U}_{lib} \mathbf{D}^m_{lib} \mathbf{W}_{lib})^{\beta^d-1}},\\
\end{split}
\end{equation}
$\beta^d$ is the value of $\beta$ used to detect the matched spectral basis matrices. Working on the tensor $\mathbf{D}^M_{lib}$, we can detect the matched spectral matrix $\mathbf{U}_{z}$ that best represents $\mathbf{U}_n$. We start by averaging along the diagonal elements of slices of the tensor $\mathbf{D}^M_{lib}$, converting the tensor into a vector $\mathbf{d}$ of size $ZK$, whose entries define the contribution of each basis vector 
\begin{equation}\label{eq:D}
\mathbf{d}= \frac{1}{M}{\sum_m \mathrm{diag}(\mathbf{D}^m_{lib})}.
\end{equation}   
The vector $\mathbf{d}$ is then normalized by its largest value in order to represent a normalized likelihood to each basis vector.} The vector $\mathbf{d}$ is divided into $Z$ sub-vectors $\mathbf{d}_z(k)$, each associated with a spectral basis matrix $\mathbf{U}_z$. {The $z$-th vector $\mathbf{d}_z(k)$ defines the contribution of each basis vector in the $z$-th matrix $\mathbf{U}_z$ to the tensor $\mathbf{\tilde{V}}^M_{\mathbf{c}_n}$}. To detect the optimal basis matrix $\mathbf{U}_{z^*}$ that best represents the \textit{n}-th source signal, the index ${z^*}$ of the optimal matrix is selected {by computing the integrated contribution of each vector $\mathbf{d}_z(k)$ as follows    
\begin{equation} \label{eq:Z}
z^*=\arg\max_z \sum_k \mathbf{d}_z(k), ~~ z=1, 2, ..., Z.   
\end{equation}
---------------------------------------------------------------------------\\
\textbf{Training}: $\mathbf{U}_{lib}$ as in sec. \ref{library}
\newline \textbf{Input}: $\mathbf{x}(\omega,l)$
\newline \textbf{Initialize}: ${\mathbf{\tilde{c}}_n}(\omega,l)$, $\mathbf{\Sigma}_{\mathbf{c}_n}(\omega,l)=I$
\newline \textbf{Iterate}: \textit{till convergence}
\newline \-~~~~Compute $\mathbf{\tilde{R}}_{\mathbf{{c}}_n}(\omega,l)$ as in (\ref{Rc})
\newline \-~~~~Build  the tensor $\mathbf{\tilde{V}}^M_{\mathbf{c}_n}$ as in sec. \ref{tensor}
\newline \-~~~~Build  the matrix $\mathbf{\tilde{V}}^{m_1m_2}_{\mathbf{c}_n}$ as in sec. \ref{tensor}
\newline \-~~~~~~\textbf{Extraction or Detection Iteration}: 
\newline \-~~~~~~~~Using $\mathbf{\tilde{V}}^M_{\mathbf{c}_n}$, Extract $\mathbf{U}_n$,  \quad $n=1, ..., N$ as in sec. \ref{sec:ESU}
\newline \-~~~~~~~~Using $\mathbf{\tilde{V}}^M_{\mathbf{c}_n}$, Detect $\mathbf{U}_n$, ~\quad $n=1, ..., N$ as in sec. \ref{detection}
\newline \-~~~~~~\textbf{Estimation Iteration}:
\newline \-~~~~~~~~~Fixing $\mathbf{U}_n$, Factorize $\mathbf{\tilde{V}}^M_{\mathbf{c}_n}$:
\newline \-~~~~~~~~~~~~ to estimate ${\theta}^{diag}_n=\{\mathbf{{W}}_n, \mathbf{V}^m_{\mathbf{R}_n}\}$ as in (\ref{eq:Wn}), (\ref{eq:Vr})
\newline \-~~~~~~~~~Fixing $\mathbf{U}_n$ and $\mathbf{{W}}_n$, Factorize $\mathbf{\tilde{V}}^{m_1m_2}_{\mathbf{c}_n}$:
\newline \-~~~~~~~~~~~~ to estimate ${\theta}^{off}_n=\{\mathbf{V}^{m_1m_2}_{\mathbf{R}_n}\}$ as in (\ref{eq:Rmm})
\newline \-~~~~\textbf{Separation}: 
\newline \-~~~~~~~Rearrange $\mathbf{V}^m_{\mathbf{R}_n}$ and $\mathbf{V}^{m_1m_2}_{\mathbf{R}_n}$ in $\mathbf{{R}}_n(\omega)$
\newline \-~~~~~~~Compute $v_n(\omega,l)=\mathbf{u}_n(\omega)\mathbf{{w}}_n(l)$
\newline \-~~~~~~~~~ $\mathbf{\Sigma}_{\mathbf{c}_n}(\omega,l)=v_n(\omega,l) \mathbf{{R}}_n(\omega)$ 
\newline \-~~~~~~~~~ $\mathbf{G}_n(\omega,l)= \mathbf{\Sigma}_{\mathbf{c}_n}(\omega,l)\mathbf{\Sigma}_{\mathbf{x}}^{-1}(\omega,l)$
\newline \-~~~~~~~~~ ${\mathbf{\tilde{c}}_n}(\omega,l)=\mathbf{G}_n(\omega,l)\mathbf{x}(\omega,l)$
\newline \-\textbf{End Iterate}
\newline \textbf{Return} 
\newline \textbf{Output}: $\mathbf{\tilde{c}}_n(\omega,l)$
\newline
\newline
$Algorithm 1$
\\---------------------------------------------------------------------------\\
Based on each tensor $\mathbf{\tilde{V}}^M_{\mathbf{c}_n}$, $n=1, ..., N$, we detect $N$ optimal spectral basis matrices. The detection phase alternates with the separation one, in order to correct a possible wrong detection occurred in the previous iteration. A wrong detection may be due to the residual from other source signals or to the coherence between the trained basis matrices. 
The proposed method is summarized in Algorithm 1, which includes an outer iteration loop and two inner iteration loops for Extraction or Detection and for Estimation respectively.

\section{EXPERIMENTAL ANALYSIS}
The experiments were carried out 
to assess the effectiveness of the proposed method
as a function of the size of spectral basis matrices $K$, the training divergence factor $\beta^t$, the detection divergence $\beta^d$, the extraction divergence $\beta^e$ and the estimation divergence $\beta^s$. Three different datasets were used for this evaluation, including simulated and live-recorded data. 
\begin{itemize}
    \item A first simulated dataset was used to analyse the detection algorithm, as well as to determine the separation performance using trained spectral basis matrices.
    \item A second live-recorded dataset was acquired in an acoustically insulated room, to evaluate the separation performance in real mixing environments when trained basis matrices are used. 
    \item The third dataset consists of simulated and live-recorded data from the SISEC2013 evaluation campaign (see http://sisec2013.wiki.irisa.fr). This was used to determine the performance  in blind and informed scenarios, when the extraction algorithm proposed in section \ref{sec:ESU} is adopted. 
    This dataset was also used to assess the  performance that can be obtained in the informed case, where the matrix $\mathbf{U}_n$ associated with each source signal is available. 
\end{itemize}
\begin{figure*}[ht]
\centering
\includegraphics[width=18cm,height=4.5cm]{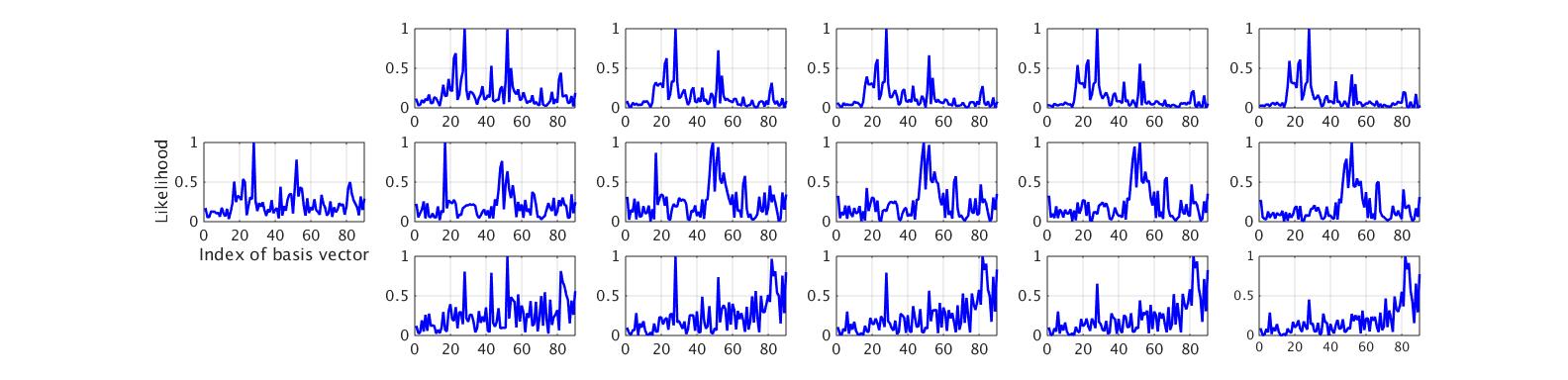}
\caption {Normalized Likelihoods $(\mathbf{d})$ of each spectral basis vector in the library $\mathbf{U}_{lib}$ as a function of the separation iterations. First column on the left for the observed mixtures, then the five columns for iterations (1, 10, 25, 40 and 65 respectively) and rows for $3$ speech source signals.}
\label{fig:detection}
\end{figure*}
\begin{figure*}[ht]
\centering
\includegraphics[width=18cm,height=4.5cm]{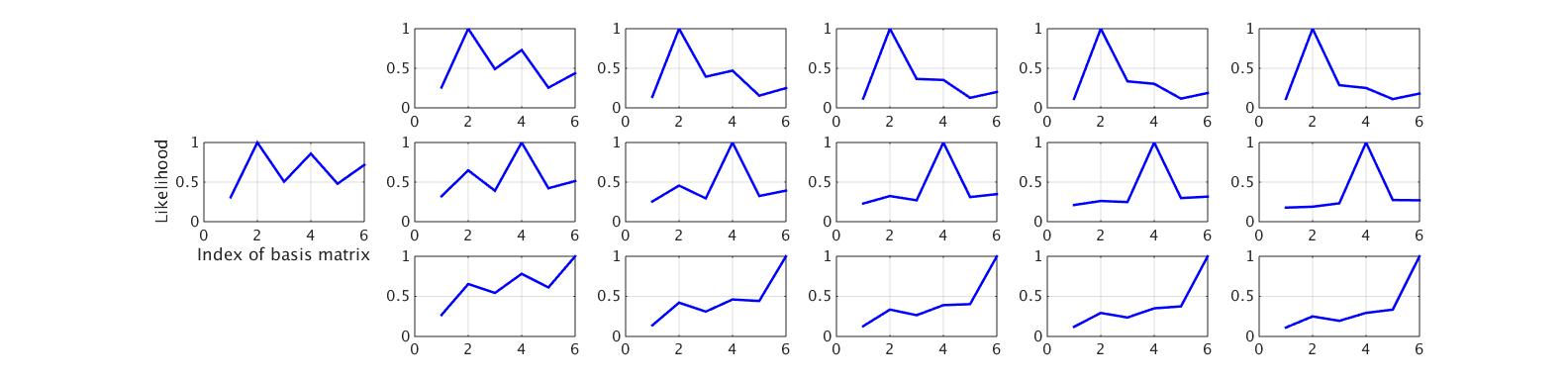}
\caption {Normalized Likelihoods of each spectral basis matrix $\mathbf{U}_z$ as a function of the separation iterations. First column on the left for the observed mixtures, then the five columns for iterations (1, 10, 25, 40 and 65 respectively) and rows for $3$ speech source signals.}
\label{fig:detection2}
\end{figure*} 
Two observed mixtures, $M=2$, and  three speech sources, $N=3$, were used for the multichannel underdetermined source separation problem. A smoothing factor $\mu=0.1$ was adopted in (\ref{eq:smooth}). The discrete time-frequency representation of the observed mixtures $\mathbf{x}(\omega,l)$ was obtained through STFT, using a Hanning analysis window with length of $128$ ms (i.e. $2048$ samples at $16$ kHz sampling rate) and shift of $64$ ms. The bi-dimensional window $\gamma$ for the computation of the empirical covariance matrix ${\mathbf{\hat{R}}}_{\mathbf{c}_n}(\omega,l)$ of the estimated source images $\mathbf{\tilde{c}}_n(\omega,l)$ was a Hanning window of size $3\times3$. Algorithm 1 was iterated till convergence, which is achieved in less than $100$ iterations. The decomposition steps were also iterated $100$ times. The separation performance was evaluated via signal-to-distortion ratio (SDR), source image-to-spatial distortion ratio (ISR), signal-to-interference ratio (SIR) and source-to-artifact ratio (SAR) expressed in decibels (dBs) \cite{Vin12}, which account for overall distortion, target distortion, residual crosstalk, and musical noise, respectively.  
\begin{table*}
\centering 
\caption{The average SDR of the simulated data as a function of $K$, $T_{60}$ and $\beta^s$, $\beta^t=0.9$.}
\label{tab:res2}
\begin{tabular}{c|c|c|c|c|c|c|c|c|c|c|c|c} 
\hline
\hline
 $T_{60}$ $(ms)$&\multicolumn{3}{|c|} {$130$}& \multicolumn{3}{|c|} {$250$}&\multicolumn{3}{|c|}{$380$}&\multicolumn{3}{|c}{Average}\\
  \hline
 $K$&15&25&40&15&25&40&15&25&40&15&25&40\\
 \hline
 $\beta^s=$ 0.9&8.53&9.18&9.04&5.50&5.61&5.32&3.58&3.46&3.23&5.87&6.08&5.86\\
 \hline
 $\beta^s=$ 0.6&9.03&9.71&9.64&6.57&6.51&5.92&4.73&4.48&4.08&6.78&6.90&6.55\\
 \hline
 $\beta^s=$ 0.3&8.64&9.26&9.66&6.60&6.85&6.84&5.02&4.95&4.44&6.75&7.02&6.98\\
 \hline
 $\beta^s=$ 0.1&7.88&8.48&9.09&6.26&6.52&6.96&4.63&4.67&4.81&6.26&6.56&6.95\\
\hline
\hline
\end{tabular}
\end{table*}
\subsection{Simulated dataset}\label{sec:Simu}
A room of size $4.45\times 3.35\times 2.5$ meters, $3$ sound sources,  and $2$ omnidirectional microphones are considered. The microphones are spaced of $0.2$ m, located in the middle of the room at the same height ($1.4$ m) as the sources. The distance between source positions and the mid-point between the microphones is randomly chosen between $0.8$ and $1.2$ m. Several source direction of arrivals (DOAs) are also simulated. The minimum angular distance between two neighboring sources is $25$ degrees, and the maximum is $40$ degrees. Synthetic room impulse responses (RIRs) are simulated through ISM \cite{ISM}, with a sampling frequency of $16$ kHz, and three reverberation times $T_{60}$, i.e., $130,~250$, or $380$ ms.} Six native Italian speakers are considered as sources, $3$ males and $3$ females. For each speaker, $20$ clean speech signals with average length of $8.75$~s were reproduced. The signals are divided into $5$ signals for testing data, and $15$ signals for training of the spectral basis matrices $\mathbf{U}_z$, $z=1,..., 6$. Six male-female combinations of mixtures were generated. This resulted in a total of $30$ test mixtures for each $T_{60}$. 
\subsubsection{Analysis of the detection algorithm}
To build the redundant library $\mathbf{U}_{lib}$ in (\ref{LIB}) for $Z=6$ spectral basis matrices, the power spectra of the training signals were computed and concatenated in the matrices $\mathbf{V}^t_z$, $z=1,2,...,6$. Each matrix was factorized with $K=15$ basis vectors, by applying NMF, as in section \ref{library}. The training divergence factor $\beta^t$ in (\ref{eq:training1}) and (\ref{eq:training2}) was assigned a value of $0.9$, quite close to the KL divergence. The \textit{z}-th spectral basis matrix $\mathbf{U}_z$ of size $1025\times 15$ was obtained, and integrated into the library of $6$ spectral basis matrices of total size $1025\times 90$. 

Let us now consider one specific case of mixing conditions, characterized by a source-to-microphone distance of $1$ m and a mixing environment with reverberation time of $250$ ms. The mixtures are generated by using two male and one female speech signals. In this specific example, the indexes $\{2, 4, 6\}$ of the basis matrices were chosen from the library and involved in the mixtures. The detection divergence factor used in (\ref{eq:detect1}) and (\ref{eq:detect2}), was $\beta^d=0.3$. Moreover, the separation divergence factor used in (\ref{eq:Wn}), (\ref{eq:Vr}) and (\ref{eq:Rmm}), was $\beta^s=0.3$. The contributions of each spectral basis vector and of spectral basis matrix were computed at each iteration of the separation phase as in (\ref{eq:D}) and (\ref{eq:Z}). In Fig. \ref{fig:detection} and Fig. \ref{fig:detection2} it can be observed that the normalized likelihood of each spectral basis vector and matrix associated with a certain target source increases while iterating the separation procedure. Therefore, the optimal index of each basis matrix becomes more identifiable with respect to the other indexes. 

The accuracy of the proposed detection algorithm depends on the value of $\beta^d$, on the configuration of the mixing process, and on the construction of the spectral basis matrices. Experiments show that the detection performance is better with small values of $\beta^d$, especially under higher reverberant conditions, and that the algorithm is more effective at low reverberation. If the mixing environment is highly reverberant, a wrong detection may happen, especially if there is much residual from other source signals, and if there is a high correlation between the trained basis matrices.

\begin{figure}
\centering
\includegraphics[scale=0.58]{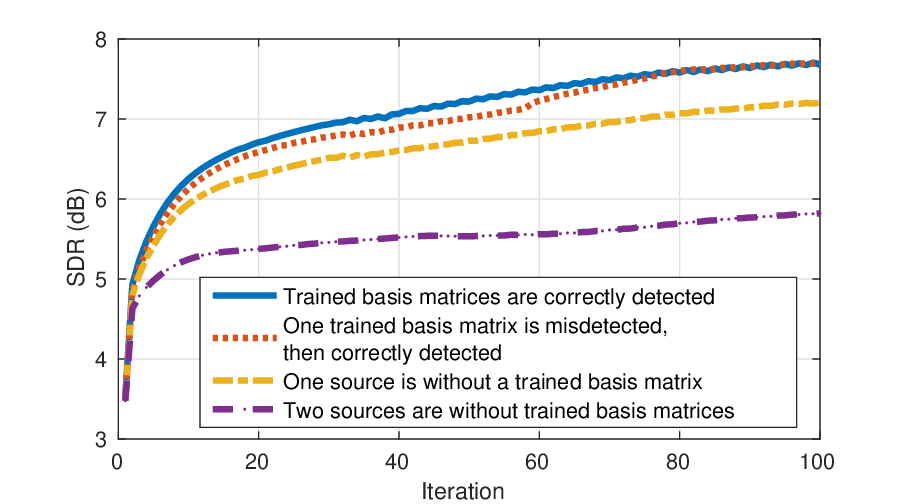}
\caption{Iterative improvement of SDR for four different detection cases.}
\label{fig:misdetection}
\end{figure}

{Fig. \ref{fig:misdetection} shows an example of the evolution of the separation performance (SDR) of the proposed method for four different cases of detection. In this experiment, three speech signals are mixed in a reverberant environment ($T_{60}=250$ ms). The distance between the sources and the midpoint between the two microphones is $1$ m, and the estimation divergence factor $\beta^s$ was $0.3$. In the first case, with a library that contains trained spectral basis matrices of all sources in observed mixtures, the detection divergence factor $\beta^d$ is $0.3$. From the first iteration the correct basis matrices are detected successfully. Furthermore, SDR increases with iterations till the convergence is achieved. The second curve concerns a case in which a correct trained basis matrix of a source is misdetected, having set $\beta^d$ to $0.6$. However, after a certain number of iterations (namely $58$), the algorithm adapts itself by detecting the correct matrix, and then it converges to the same SDR as in the previous case. Although a correct trained basis matrix is initially misdetected, a better separation is achieved than in a case where the library does not contain the correct matrix. The third curve refers to the latter situation. The best the algorithm can do in this case is to select the matrix having the highest coherence with observations. The fourth curve represents the case in which two sources are not seen during the training.

\subsubsection{Source separation} 
Table \ref{tab:res2} describes the average SDR as a function of $K$, $T_{60}$ and $\beta^s$, with fixed $\beta^t=0.9$. With a $T_{60}=130$ ms, the best separation performance is obtained when moderate values (between $0.3$ and $0.6$) are assigned to $\beta^s$. As the environment becomes more reverberant ($T_{60}=250$ or $380$ ms), the observed mixtures tend to overlap more and more. To improve the estimation of the parameters in this environment, we need to push up the factorization sparsity by using smaller values of $\beta^s$, which is confirmed by the better performance obtained with smaller values of $\beta^s$ (between $0.3$ and $0.1$). On the other side, the proposed method achieves better results with trained basis matrices of large size ($K\geq 25$) in environments with low reverberation ($T_{60}=130$ ms), while the opposite is observed in a more reverberant environment ($T_{60}=380$ ms). On the average, a small value of $\beta^s$ ($0.1$) performs better than a large one ($0.9$). Moreover, the best performance is obtained when $\beta^s$ is in the range between $0.3$ and $0.6$, but this range also depends on the value of $\beta^t$.} In general, the experiments show the importance of choosing proper values of $\beta$ for training the basis matrix $\mathbf{U}_n$ and for estimating the model parameters $\theta$ that guarantee a good separation performance.

\subsection{Live-recorded dataset}
This dataset was recorded in a room of size $6.5\times 3\times 2.2$ m, and using $2$ omnidirectional microphones spaced of $16$ $cm$. The microphones are located close to the center of one of the room walls and are at the same height ($1.5$ m) as the $3$ speakers. The distance between source positions and the mid-point between the microphones is $1.5$ m, with DOAs at $55$, $90$ and $125$ degrees. The measured reverberation time of the room is around $220$ ms. The mixtures were recorded at a sampling frequency of $16$ kHz. In these experiments, we used the same speech source signals and trained spectral basis matrices that were adopted in the simulated scenario described in section \ref{sec:Simu}. Table \ref{tab:shine} shows the corresponding results as a function of $K$ and $\beta^s$, with fixed $\beta^t=0.9$. The performance is close to what obtained in the simulated experiments at $T_{60}=250$ ms. The best performance is obtained when $K$ is  large enough (between $20$ and $40$), and $\beta^s$ is in the lower range. 
\begin{table}
\centering 
\caption{The average SDR of the live-recorded data as a function of $K$ and $\beta^s$, $\beta^t=0.9$.}
\label{tab:shine}
\begin{tabular}{c|c|c|c|c} 
\hline
\hline
 $K$&$10$&$20$&$30$&$40$\\
 \hline
 $\beta^s=$ 0.9&4.52&4.82&5.10&5.43\\
 \hline
 $\beta^s=$ 0.6&5.60&5.70&5.79&5.88\\
 \hline
 $\beta^s=$ 0.3&5.72&6.62&6.78&6.62\\
 \hline
 $\beta^s=$ 0.1&5.67&6.46&6.42&6.43\\
\hline
\hline
\end{tabular}
\end{table}

\subsection{Dataset of SISEC2013} 
The development dataset $dev1$ of SISEC2013 (under-determined speech and music mixtures) was used for further evaluation of the proposed method. The dataset consists of $4$ synthetic convolutive and $4$ live-recorded stereo mixtures of three Japanese and English speakers. All the mixtures are of length $10$ s and sampled at $16$ kHz. The synthetic convolutive filters are generated with the Roomsim toolbox \cite{Roomsim}. They simulate $2$ omnidirectional microphones placed $1$ m apart in a room of dimension $4.45\times 3.35\times 2.5$ m with reverberation times of $130$ and $250$ ms. The distance between source positions and the mid-point between the microphones varies between $0.8$ and $1.2$ m. For all the mixtures the DOAs vary between $60$ and $300$ angular degrees, with a minimum spacing of $15$ degrees. This dataset was used to evaluate the proposed method both in a {\textit{blind scenario}}, where the extraction algorithm of the spectral basis matrices $\mathbf{U}_n$, described in section \ref{sec:ESU} is adopted, and in an {\textit{informed scenario}}, where the true spectral basis matrices $\mathbf{U}_n$ are trained as in section \ref{library}. 
\begin{table}
\centering 
\caption{The performance of informed and blind source separation of 3-Females of live-recorded mixtures from the SISEC dataset.}
\label{tab:dev1}
\begin{tabular}{c|c|c|c|c|c|c} 
\hline
\hline
  $\mathbf{U}_n$&\multicolumn{3}{|c|}{Informed}&\multicolumn{3}{|c}{Blind}\\
  \hline
 Div. Factors &\multicolumn{3}{|c|}{$\beta^t=0.9,\beta^s=0.3$}&\multicolumn{3}{|c}{$\beta^e=0.6,\beta^s=0.6$}\\
  \hline
 Source images&$\mathbf{\tilde{c}}_1$&$\mathbf{\tilde{c}}_2$&$\mathbf{\tilde{c}}_3$&$\mathbf{\tilde{c}}_1$&$\mathbf{\tilde{c}}_2$&$\mathbf{\tilde{c}}_3$\\
 \hline
 SDR&11.27&10.42&10.22&9.63&8.67&9.17\\
 \hline
 ISR&15.13&14.56&17.76&12.90&12.82&16.50\\
 \hline
 SIR&17.26&18.02&13.12&15.28&13.97&11.95\\
 \hline
 SAR&14.99&12.99&13.68&14.32&12.33&12.76\\
\hline
\hline
\end{tabular}
\end{table}
\begin{table*}
\centering 
\caption{The SDR of informed separation of live-recorded stereo mixtures from SISEC, $\beta^t=0.9$.}
\label{tab:inf}
\begin{tabular}{c||c|c|c|c|c|c||c|c|c|c|c|c||c|c|c} 
\hline
\hline
$T_{60}$ &\multicolumn{6}{|c||}{$130~\textrm{ms}$}&\multicolumn{6}{|c||}{$250~\textrm{ms}$}&\multicolumn{3}{c}{\multirow{2}{*}{Average}}\\
\cline{1-13}
Source&\multicolumn{3}{|c|}{Three females}&\multicolumn{3}{|c||}{Three males}&\multicolumn{3}{|c|}{Three females}&\multicolumn{3}{|c||}{Three males}\\
\hline
 $K$ &$20$&$35$&$50$&$20$&$35$&$50$&$20$&$35$&$50$&$20$&$35$&$50$&$20$&$35$&$50$\\
\hline
 $\beta^s=$ 0.9&9.45&9.88&9.90&8.43&8.83&8.95&6.34&4.21&4.14&7.70&7.87&7.53&7.98&7.70&7.63\\
 \hline
 $\beta^s=$ 0.6&10.08&10.20&10.07&8.57&8.90&9.10&6.72&6.11&5.93&7.77&7.97&8.25&8.29&8.30&8.34\\
 \hline
 $\beta^s=$ 0.3&9.80&10.70&10.56&8.07&8.60&8.94&9.05&9.40&9.01&7.39&7.82&8.11&8.58&9.13&9.16\\
 \hline
 $\beta^s=$ 0.1&8.95&9.74&10.18&7.63&8.15&8.50&7.88&8.16&8.86&6.76&7.33&7.78&7.81&8.35&8.83\\
\hline
\hline
\end{tabular}
\end{table*}

\begin{table*}
\centering 
\caption{The average SDR of blind separation of stereo mixtures from SISEC, $T_{60}=130~\textrm{ms}$.}
\label{tab:bss}
\begin{tabular}{c|c|c|c|c|c|c|c|c|c|c|c|c|c|c|c|c|c|c} 
\hline
\hline
Env. &\multicolumn{9}{|c|}{Synthetic convolutive mixtures}&\multicolumn{9}{|c}{Live-recorded mixtures}\\
\hline
 $K$ &\multicolumn{3}{|c|} {$20$}&\multicolumn{3}{|c|} {$35$}&\multicolumn{3}{|c|} {$50$}&\multicolumn{3}{|c|} {$20$}&\multicolumn{3}{|c|} {$35$}&\multicolumn{3}{|c} {$50$}\\
  \hline
 $\beta^e$&1.2&0.9&0.6&1.2&0.9&0.6&1.2&0.9&0.6&1.2&0.9&0.6&1.2&0.9&0.6&1.2&0.9&0.6\\
 \hline
 $\beta^s=$ 0.9&5.53&5.97&6.00&5.70&5.79&5.63&5.78&5.80&5.68&7.46&7.54&5.83&7.74&7.58&7.95&7.55&7.62&7.80\\
 \hline
 $\beta^s=$ 0.6&5.92&6.32&6.49&5.71&6.22&5.97&6.18&6.17&5.76&7.49&7.27&6.98&7.66&7.48&7.76&7.61&7.90&$\mathbf{8.16}$\\
 \hline
 $\beta^s=$ 0.3&5.51&6.79&6.75&5.66&6.01&6.81&5.71&6.90&6.08&6.07&7.06&6.76&6.31&7.20&7.31&7.23&7.62&7.98\\
 \hline
 $\beta^s=$ 0.1&5.07&5.55&5.75&5.44&6.50&$\mathbf{7.07}$&5.34&6.36&6.20&5.26&5.59&5.41&5.48&6.76&6.74&6.26&7.10&7.11\\
\hline
\hline
\end{tabular}
\end{table*}

\begin{table*}
\centering 
\caption{The SDR of blind separation of live-recorded stereo mixtures from SISEC, $T_{60}=130$ \textrm{ms}.}
\label{tab:bss1}
\begin{tabular}{c||c|c|c|c|c|c|c|c|c||c|c|c|c|c|c|c|c|c} 
\hline
\hline
Source &\multicolumn{9}{|c||}{Three females}&\multicolumn{9}{|c}{Three males}\\
\hline
 $K$ &\multicolumn{3}{|c|} {$20$}&\multicolumn{3}{|c|} {$35$}&\multicolumn{3}{|c||} {$50$}&\multicolumn{3}{|c|} {$20$}&\multicolumn{3}{|c|} {$35$}&\multicolumn{3}{|c} {$50$}\\
  \hline
 $\beta^e$&1.2&0.9&0.6&1.2&0.9&0.6&1.2&0.9&0.6&1.2&0.9&0.6&1.2&0.9&0.6&1.2&0.9&0.6\\
 \hline
 $\beta^s=$ 0.9&8.10&8.27&5.98&8.70&8.35&8.83&8.48&8.80&9.04&6.82&6.81&5.68&6.78&6.81&7.07&6.62&6.44&6.56\\
 \hline
 $\beta^s=$ 0.6&7.75&8.35&8.12&8.52&8.70&8.72&8.72&8.99&$\mathbf{9.20}$&7.23&6.19&5.84&6.80&6.26&6.80&6.50&6.81&$\mathbf{7.13}$\\
 \hline
 $\beta^s=$ 0.3&7.29&7.51&7.72&7.40&8.14&8.38&8.43&8.38&9.11&4.85&6.61&5.80&6.00&6.26&6.24&6.03&6.86&6.85\\
 \hline
 $\beta^s=$ 0.1&5.66&6.97&6.60&5.80&7.26&8.30&7.12&7.99&8.85&4.86&4.21&4.22&5.16&6.26&5.17&5.40&6.21&5.37\\
\hline
\hline
\end{tabular}
\end{table*}

\begin{table*}
\centering 
\caption{The SDR of blind separation of live-recorded stereo mixtures from SISEC, $T_{60}=250$ ms.}
\label{tab:bss2}
\begin{tabular}{c||c|c|c|c|c|c|c|c|c||c|c|c|c|c|c|c|c|c} 
\hline
\hline
Source &\multicolumn{9}{|c||}{Three females}&\multicolumn{9}{|c}{Three males}\\
\hline
 $K$ &\multicolumn{3}{|c|} {$20$}&\multicolumn{3}{|c|} {$35$}&\multicolumn{3}{|c||} {$50$}&\multicolumn{3}{|c|} {$20$}&\multicolumn{3}{|c|} {$35$}&\multicolumn{3}{|c} {$50$}\\
  \hline
 $\beta^e$&1.2&0.9&0.6&1.2&0.9&0.6&1.2&0.9&0.6&1.2&0.9&0.6&1.2&0.9&0.6&1.2&0.9&0.6\\
 \hline
 $\beta^s=$ 0.9&3.65&3.42&3.71&3.55&3.63&3.24&3.56&3.49&3.36&3.90&3.32&3.66&4.71&3.17&3.44&4.03&3.96&3.46\\
 \hline
 $\beta^s=$ 0.6&3.81&4.52&4.01&4.54&4.65&3.45&5.60&3.54&3.54&3.47&3.48&3.52&5.04&4.01&3.69&4.09&4.22&4.10\\
 \hline
 $\beta^s=$ 0.3&3.59&6.44&6.29&4.55&4.95&$\mathbf{6.50}$&3.67&4.66&4.36&3.59&3.74&4.20&4.48&$\mathbf{5.17}$&4.34&4.87&5.07&4.47\\
 \hline
 $\beta^s=$ 0.1&3.63&5.74&6.19&4.61&5.86&$\mathbf{6.50}$&4.67&5.42&4.88&3.37&3.40&3.58&4.85&4.33&4.31&5.11&5.00&4.80\\
\hline
\hline
\end{tabular}
\end{table*}

\begin{table*}
\centering 
\caption{Performance comparison of live-recorded stereo mixtures from SISEC, $T_{60}=130$ {ms}.}
\label{tab:bssc1}
\begin{tabular}{c||c|c|c|c|c||c|c|c|c|c} 
\hline
\hline
Source&\multicolumn{5}{c||}{Three females}&\multicolumn{5}{c}{Three males}\\
\hline
\multirow{1}{*}{Method}&\multicolumn{3}{c|}{Proposed}&\multirow{5}{*}{Nesta}&\multirow{5}{*}{Cho}&\multicolumn{3}{c|}{Proposed}&\multirow{5}{*}{Nesta}&\multirow{5}{*}{Cho}\\ 
\cline{1-4} \cline{7-9} $\mathbf{U}_n$&Inform&Blind&\multirow{4}{*}{Initialize}&&&Inform&Blind&\multirow{4}{*}{Initialize}&&\\
\cline{1-3} \cline{7-8} $K$&50&50&&&&50&50&&&\\
\cline{1-3} \cline{7-8} $\beta^t$/$\beta^e$&0.9&0.6&&\cite{Nesta}&\cite{cho}&0.9&0.6&&\cite{Nesta}&\cite{cho}\\
\cline{1-3} \cline{7-8} $\beta^s$&0.6&0.6&&&&0.6&0.6&&&\\
\hline
SDR&10.07&9.20&3.92&7.70&8.40&9.10&7.13&2.58&6.50&6.50\\
\hline
ISR&15.14&14.10&9.04&10.50&13.00&14.10&12.06&7.37&9.30&11.40\\
\hline
SIR&14.83&13.70&8.11&13.30&12.60&13.80&11.43&5.18&10.90&10.00\\
\hline
SAR&13.66&13.10&6.92&11.80&12.10&12.20&10.15&5.06&9.60&10.50\\
\hline
\hline
\end{tabular}
\end{table*}

\begin{table*}
\centering 
\caption{Performance comparison of live-recorded stereo mixtures from SISEC, $T_{60}=250$ ms.}
\label{tab:bssc2}
\begin{tabular}{c||c|c|c|c|c||c|c|c|c|c} 
\hline
\hline
Source&\multicolumn{5}{c||}{Three females}&\multicolumn{5}{c}{Three males}\\
\hline
\multirow{1}{*}{Method}&\multicolumn{3}{c|}{Proposed}&\multirow{5}{*}{Nesta}&\multirow{5}{*}{Cho}&\multicolumn{3}{c|}{Proposed}&\multirow{5}{*}{Nesta}&\multirow{5}{*}{Cho}\\ 
\cline{1-4} \cline{7-9} $\mathbf{U}_n$&Inform&Blind&\multirow{4}{*}{Initialize}&&&Inform&Blind&\multirow{4}{*}{Initialize}&&\\
\cline{1-3} \cline{7-8} $K$&35&35&&&&35&35&&&\\
\cline{1-3} \cline{7-8} $\beta^t$/$\beta^e$&0.9&0.6&&\cite{Nesta}&\cite{cho}&0.9&0.6&&\cite{Nesta}&\cite{cho}\\
\cline{1-3} \cline{7-8} $\beta^s$&0.3&0.3&&&&0.3&0.3&&&\\
\hline
SDR&9.40&6.50&2.32&6.00&6.10&7.82&4.34&1.95&5.20&6.00\\
\hline
ISR&14.14&11.08&7.08&8.90&10.90&12.71&8.90&6.40&8.30&10.50\\
\hline
SIR&14.10&10.16&4.40&10.60&9.00&12.36&7.71&3.76&9.00&9.10\\
\hline
SAR&12.40&10.61&5.95&8.70&10.00&10.44&7.38&4.70&8.00&9.20\\
\hline
\hline
\end{tabular}
\end{table*}

\begin{figure}
\centering
\includegraphics[width=8.8cm,height=3.7cm]{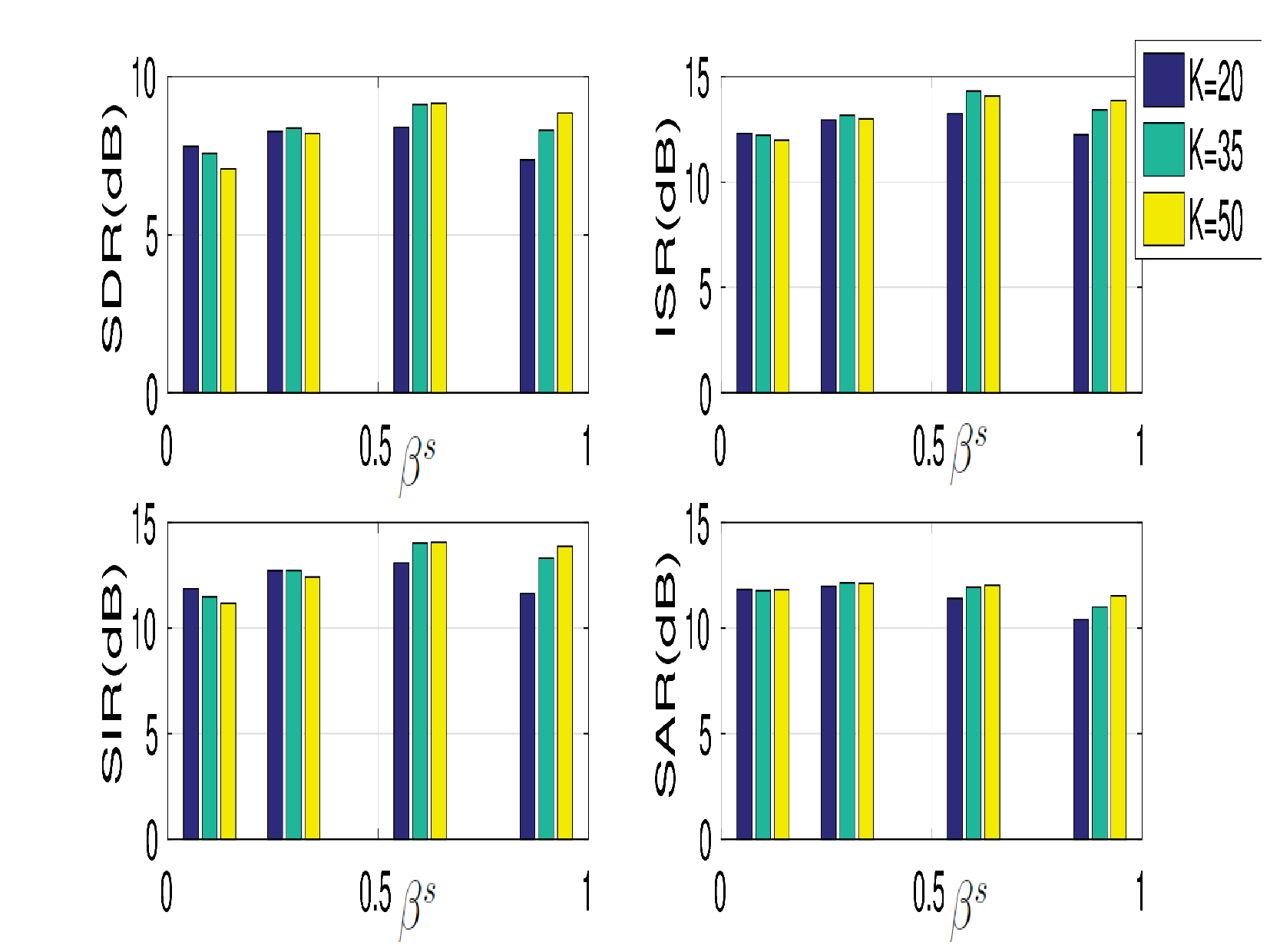}
\caption{The average separation performance of the informed case of the SISEC dataset, as a function of $K$ and $\beta^s$ ($\beta^t=0.9$). The horizontal axis indicates the values of the tested $\beta^s=[0.1~0.3~0.6~0.9]$.}
\label{fig:sisec1}
\end{figure}

Table \ref{tab:dev1} reports different results concerning both informed and blind scenarios. Only the live-recorded stereo mixture of 3-female speech signals was used in these experiments, with a reverberation time equal to $130$ ms. The spectral basis matrix $\mathbf{U}_n$ is either trained on power spectra of sources with $K=35$, or extracted with $K=50$, i.e. the best performing sizes of $\mathbf{U}_n$. Comparing informed and blind scenarios, an average decrease of $1.5$ dBs SDR is observed in the latter case.

\subsubsection{{Informed source separation}}
The basis matrices $\mathbf{U}_n$ were trained, using power spectra of sources in the mixtures, with different sizes $K=[20~35~50]$ and $\beta^t=0.9$. The selected values of $\beta^s$ were again $0.1,~0.3,~0.6,$ and $0.9$. Fig. \ref{fig:sisec1} shows the average separation performance of all the $8$ mixtures, for all values of $K$. The best performance is obtained when the value of $\beta^s$ equals to $0.6$. Table \ref{tab:inf} reports the detailed SDR of the informed case for each live-recorded mixture. The results confirm that lower values of $\beta^s$ are preferable in more reverberant environments. This is more evident for mixtures of female voices that, being in the average more sparse, emphasize this trend.

\subsubsection{Blind source separation}
We evaluated the performance in a blind scenario using $4$ observed mixtures ($2$ synthetic convolutive and $2$ live-recorded, all with reverberation time of $130$ ms). The spectral basis matrices $\mathbf{U}_n$ are extracted as in section \ref{sec:ESU}. Table \ref{tab:bss} reports the average SDR for different settings of the number of spectral bases $K$, the extraction divergence factor $\beta^e$ and the estimation divergence factor $\beta^s$ (bold numbers indicate outstanding performance). In the case of synthetic convolutive mixtures, the best average SDR ($7.07$ dBs) is achieved when $K=35$, $\beta^e=0.6$ and $\beta^s=0.1$, which is still in line with the trend observed in the previous sections. 

On the other hand, in the experiments with live-recorded mixtures the best performance is obtained when $K=50$, $\beta^e$ between $0.6$ and $0.9$, and $\beta^s$ in its moderate range (between $0.3$ and $0.6$). Overall, this appears similar to the trend observed in the informed case of the live-recorded mixtures. The contrasting experimental evidence observed moving from synthetic convolutive mixtures to live-recorded ones might be explained with a higher reverberation time than 130 ms characterizing synthetic recordings.

\subsubsection{Performance comparison}
For comparison purposes, Tables \ref{tab:bss1} and \ref{tab:bss2} report the detailed SDR of each mixture of the live-recorded mixtures, reported in the right part of Table \ref{tab:bss}. For mixtures of female speech signals in both cases of mixing environments, $\beta^e=0.6$ performs the best with $K=50$, in case of $T_{60}=130$ ms, and $K=35$, in case of $T_{60}=250$ ms. The results with female voices again outperform those obtained with male voices, and smaller values of $\beta^s$ are more suitable in the case of $T_{60}=250$ ms. This seems to be a relevant experimental evidence for a challenging situation related to a more reverberant condition.
 
The proposed algorithm was compared to the blind source separation algorithms in \cite{Nesta,cho}. Table \ref{tab:bssc1} and \ref{tab:bssc2} report the results obtained with different evaluation metrics. The experiments were conducted after having explored different choices for $K$,  $\beta^t$, $\beta^e$, $\beta^s$, in order to individuate the optimal setting for each reverberation time. The results obtained with these optimal settings show that both $Informed$ and $Blind$ definitely outperform the $Initialize$ case (i.e., the case in which we adopt the binary clustering initialization described in Section
\ref{S:3}). Furthermore, in three of the four cases our proposed method outperforms the other two algorithms. In particular, the worse performance obtained in the case of male mixtures under the higher reverberation time indicates that our method is less effective when observations are less sparse.

\section{CONCLUSION}
In this work, we tackled the problem of underdetermined audio source separation in reverberant environments. The proposed work adopts local Gaussian modeling of the mixing process. The paper describes a new estimation algorithm of the parameters of the model by applying nonnegative tensor/matrix factorization, given source-based prior information. Following this direction, the parameters are jointly estimated, and the related artifacts (e.g., residuals due to source crosstalk) are consequently reduced. To perform the estimation, spectral basis matrices of power spectra of source signals in observed mixtures are assumed to be available as prior information. In a separate step, the basis matrices are either extracted or detected. A new method is proposed, which uses nonnegative tensor factorization for extracting the matrices. In this case, the algorithm fully works in a blind scenario. However, to obtain a better separation performance, in the other case the matrices are made indirectly available through a pre-trained redundant library of spectral basis matrices. Furthermore, we propose a new method based on  nonnegative tensor factorization, to detect the basis matrices that best represent the power spectra of the source signals in the observed mixtures. 

For each of the training, detection, extraction and estimation phases, the factorization is performed using the $\beta$-divergence and by applying the widely used multiplicative update rules. The sparsity of factorization can be controlled by tuning the value of $\beta$, this way minimizing residual artifacts. This is an important issue as speech signals are sparse in their nature.
Accordingly, we tested several values of $\beta$ for each task in order to identify the best performing ones. We found that the best choice of $\beta$ to estimate the parameters, is in the interval between $0.3$ and $0.6$, in both cases of extracting the basis matrices and of detecting the trained ones. On the other hand, to extract the basis matrices, we found that the best choice of $\beta$ for mixtures of female voices is between $0.6$ and $0.9$. However large values ($>0.9$) perform better for mixtures of male voices. In terms of size of the basis matrices, large size matrices perform better than small ones in mixing environments with low reverberation, while the opposite holds in mixing environments with high reverberation. The experimental results show that the algorithm can work with approximately the same efficiency in both cases of simulated and real environments. 

Experiments also showed that the proposed algorithm outperforms two of the recently proposed blind source separation algorithms.  
An average SDR improvement of $2.6$ dB is achieved in the informed cases. In the blind scenario, the algorithm outperforms the other algorithms in mixing environments with low reverberation ($T_{60}=130$ $ms$), and in both cases of mixtures of female and male voices. In mixing environments with moderate reverberation ($T_{60}=250$ $ms$), the algorithm performs better in the case of female voices but it fails in the case of male voices. 

In the future, we plan to improve the proposed algorithm to get better performance in mixing environments with high reverberation, by investigating on efficient factorization algorithms that are more robust against the reverberation. 
We also plan to work on self adaptation of the divergence factor $\beta$ as a function of the signal levels in the observed mixtures.

\appendices
\section{the MU rule for the $\beta$-divergence} \label{appendix}
The divergence in (\ref{eq:betadiv}), between the element $a$ and its decomposition $bcd$, can de defined as    
\begin{align*}
d_{\beta}(a|bcd)= \frac{a^{\beta}+(\beta-1)(bcd)^{\beta}-\beta a(bcd)^{\beta-1}}{\beta(\beta-1)}.
\end{align*}
For example to minimize the divergence with respect to the element $b$, the partial derivative of $d_{\beta}(a/bcd)$ is computed as
\begin{align*}
g= \frac{\beta(\beta-1)cd(bcd)^{\beta-1}-\beta ({\beta-1})acd(bcd)^{\beta-2}}{\beta(\beta-1)}.
\end{align*}
The positive part of the derivative is defined as 
\begin{align*}
 g_+=cd(bcd)^{\beta-1},
 \end{align*}
and the negative part is represented as
\begin{align*}
g_-=acd(bcd)^{\beta-2}.  
 \end{align*}  
The MU rule to update the element $b$ is described in terms of $g_+$ and $g_-$ as
\begin{align*}
b\leftarrow b \frac{g_-}{g_+}=b \frac{acd(bcd)^{\beta-2}}{cd(bcd)^{\beta-1}}.
\end{align*}   
Furthermore, we could define the update rule using the MU for each element. Furthemore, this element-wise update rule can be easily extended for matrix factorization, respecting the dimensions.   

For tensor factorization, let us assume that $\mathbf{A}^M$ is a tensor of size $1 \times 1 \times M$. By decomposing the tensor into two elements $b$ and $d$, and a tensor $\mathbf{C}^M$, the divergence can be described as 
\begin{align*}
\begin{split}
\sum_m d_{\beta}(a^m|bc^md)= \frac{1}{\beta(\beta-1)}\sum_m&(a^m)^{\beta}+(\beta-1)(bc^md)^{\beta}\\
&-\beta a^m(bc^md)^{\beta-1},
\end{split}
\end{align*}
where $a^m$ and $c^m$ are the \textit{m}-th elements of $\mathbf{A}^M$ and $\mathbf{C}^M$, respectively. 
To minimize the divergence with respect to, for example, the element $b$, the partial derivative $g$ is computed
\begin{align*}
 g= \sum_m {c^md(bc^md)^{\beta-1}-\sum_m a^mc^md(bc^md)^{\beta-2}}.
\end{align*}
Accordingly, the MU rule to update the element $b$ is given by
\begin{align*}
b\leftarrow b \frac{\sum_m a^mc^md(bc^md)^{\beta-2}}{\sum_m {c^md(bc^md)^{\beta-1}}}.
\end{align*}   
Respecting the dimensions, this update rule can be easily extended to decompose a tensor into tensors and matrices.

\vspace{-0.6cm}

  \begin{IEEEbiography}
{Mahmoud Fakhry}
  received the B.Sc. and M.Sc. degrees in Electrical Engineering from the Department of Electrical Engineering, Aswan University, Egypt and the PhD degree in Information and Communication Technology from the University of Trento, Italy, in 2016. From October 2012 to November 2016, he was with the Center for Information and Communication Technology, Fondazione Bruno Kessler, Trento, Italy, as PhD student. Since December 2016, he is a postdoc researcher at the Audio Analysis Lab, Aalborg University, Denmark. His research interests include source modeling using nonnegative matrix and tensor factorization, analysis of dynamic systems, speech separation and enhancement, and voice analysis for early diagnosis of diseases.
\end{IEEEbiography} 

\vspace{-1cm}

  \begin{IEEEbiographynophoto}
  {Piergiorgio Svaizer}
  received the degree in electronic engineering (communications) from the University of Padua, Padua, Italy, in 1989. He is a senior researcher at the Center for Information and Communication Technology, Fondazione Bruno Kessler, Trento, Italy. His main research interests include speech analysis, enhancement and separation, front-end processing and noise reduction algorithms for speech recognition in adverse conditions, and microphone arrays for acoustic scene analysis and acoustic surveillance.
\end{IEEEbiographynophoto}  
    
\vspace{-1cm}    
    
  \begin{IEEEbiographynophoto}
  {Maurizio Omologo}
(M’88) received the “Laurea” degree (with
honors) in electrical engineering from the University
of Padua in 1984.
He is the head of the SHINE
(Speech-Acoustic Scene Analysis and Interpretation)
research unit of the Center for Information and Communication Technology, Fondazione Bruno Kessler, Trento, Italy. 
His current research
interests include audio and speech processing, acoustic scene analysis,
and automatic speech recognition, in particular for distant-talking scenarios.
He is author of more than 100 papers in major international conferences
and journals in the field.
\end{IEEEbiographynophoto}  

\end{document}